\documentclass[fleqn,usenatbib]{mnras}

\usepackage{newtxtext,newtxmath}

\usepackage[T1]{fontenc}

\usepackage{graphicx}	
\usepackage{amsmath}	
\usepackage{color}
\usepackage{soul}
\usepackage{hyperref}
\usepackage{xspace}
\usepackage{CJKutf8}
\usepackage{deluxetable}
\usepackage{afterpage}

\hypersetup{    
  colorlinks      = {true},
  linkcolor       = {blue},
  citecolor       = {blue},
  urlcolor        = {blue},
}

\newcommand{\code}[1]{\texttt{#1}\xspace}

\newcommand{\lp}{\texttt{LESSPayne}\xspace}

\newcommand{\unit}[1]{\ensuremath{\mathrm{\,#1}}\xspace}

\newcommand{\Teff}{\ensuremath{T_\mathrm{eff}}\xspace}
\newcommand{\logg}{\ensuremath{\log\,g}\xspace}

\newcommand{\vt}{\ensuremath{v_\mathrm{t}}\xspace}

\newcommand{\kms}{\unit{km\,s^{-1}}}

\title[Abundances of Wukong/LMS-1]{Extending the Chemical Reach of the H3 Survey: Detailed Abundances of the Dwarf-galaxy Stellar Stream Wukong/LMS-1\thanks{This paper includes data gathered with the 6.5~meter Magellan Telescopes located at Las Campanas Observatory, Chile.}}

\author[G. Limberg et al.]{Guilherme Limberg,$^{1,2,3}$\thanks{E-mail: guilherme.limberg@usp.br}
Alexander P. Ji,$^{2,3}$
Rohan P. Naidu,$^{4}$
Anirudh Chiti,$^{2,3}$
Silvia Rossi,$^{1}$ 
\newauthor
Sam A. Usman,$^{2,3}$
Yuan-Sen Ting (\begin{CJK*}{UTF8}{gkai}丁源森\end{CJK*}),$^{5,6,7}$
Dennis Zaritsky,$^{8}$
Ana Bonaca,$^{9}$
Lais Borbolato,$^{1}$
\newauthor
Joshua S. Speagle (\begin{CJK*}{UTF8}{gkai}沈佳士\end{CJK*}),$^{10,11,12,13}$
Vedant Chandra,$^{14}$
Charlie Conroy$^{14}$ \smallskip
\\ \\
$^1$ Universidade de S\~ao Paulo, IAG, Departamento de Astronomia, SP 05508-090, S\~ao Paulo, Brazil\\
$^{2}$ Department of Astronomy \& Astrophysics, University of Chicago, 5640 S Ellis Avenue, Chicago, IL 60637, USA\\
$^{3}$ Kavli Institute for Cosmological Physics, University of Chicago, Chicago, IL 60637, USA\\
$^{4}$ Department of Physics and Kavli Institute for Astrophysics and Space Research, Massachusetts Institute of Technology, Cambridge, MA 02139, USA\\
$^{5}$ Research School of Astronomy \& Astrophysics, Australian National University, Weston, ACT 2611, Canberra, Australia\\
$^{6}$ School of Computing, Australian National University, Acton, ACT 2601, Canberra, Australia\\
$^{7}$ Department of Astronomy, The Ohio State University, 140 West 18th Avenue, Columbus, OH 43210, USA\\
$^{8}$ Steward Observatory, University of Arizona, 933 N. Cherry Ave., Tucson, AZ 85721-0065, USA \\
$^{9}$ Observatories of the Carnegie Institution for Science, 813 Santa Barbara Street, Pasadena, CA 91101, USA\\
$^{10}$ Department of Statistical Sciences, University of Toronto, Toronto, ON M5G 1Z5, Canada\\
$^{11}$ David A. Dunlap Department of Astronomy \& Astrophysics, University of Toronto, Toronto, ON M5S 3H4, Canada\\
$^{12}$ Dunlap Institute for Astronomy \& Astrophysics, University of Toronto, Toronto, ON M5S 3H4, Canada\\
$^{13}$ Data Sciences Institute, University of Toronto, Toronto, ON M5G 1Z5, Canada\\
$^{14}$ Center for Astrophysics $\mid$ Harvard \& Smithsonian, 60 Garden Street, Cambridge, MA 02138, USA \vspace{-5mm}
}

\date{Accepted XXX. Received YYY; in original form ZZZ  \vspace{-5mm}}

\pubyear{2023}

\begin{document}
\label{firstpage}
\pagerange{\pageref{firstpage}--\pageref{lastpage}}
\maketitle

\begin{abstract}
We present the first detailed chemical-abundance analysis of stars from the dwarf-galaxy stellar stream Wukong/LMS-1 covering a wide metallicity range ($-3.5 < \rm[Fe/H] \lesssim -1.3$). We find abundance patterns that are effectively indistinguishable from the bulk of Indus and Jhelum, a pair of smaller stellar streams proposed to be dynamically associated with Wukong/LMS-1. We confirmed a carbon-enhanced metal-poor star ($\rm[C/Fe] > +0.7$ and $\rm[Fe/H] \sim -2.9$) in Wukong/LMS-1 with strong enhancements in Sr, Y, and Zr, which is peculiar given its solar-level [Ba/Fe]. Wukong/LMS-1 stars have high abundances of $\alpha$ elements up to $\rm[Fe/H] \gtrsim -2$, which is expected for relatively massive dwarfs.
Towards the high-metallicity end, Wukong/LMS-1 becomes $\alpha$-poor, revealing that it probably experienced fairly standard chemical evolution. We identified a pair of N- and Na-rich stars in Wukong/LMS-1, reminiscent of multiple stellar populations in globular clusters. This indicates that this dwarf galaxy contained at least one globular cluster that was completely disrupted in addition to two intact ones previously known to be associated with Wukong/LMS-1, which is possibly connected to similar evidence found in Indus. From these $\geq$3 globular clusters, we estimate the total mass of Wukong/LMS-1 to be ${\approx}10^{10} M_\odot$, representing ${\sim}1$\% of the present-day Milky Way. Finally, the [Eu/Mg] ratio in Wukong/LMS-1 continuously increases with metallicity, making this the first example of a dwarf galaxy where the production of $r$-process elements is clearly dominated by delayed sources, presumably neutron-star mergers.
\end{abstract}

\begin{keywords}
stars: abundances -- Galaxy: halo -- Galaxy: kinematics and dynamics -- galaxies: Local Group  \vspace{-5mm} 
\end{keywords}



\section{Introduction}

\indent\indent In a cold dark matter-dominated cosmology, massive Milky Way-like halos are built up through successive accretion of dwarf galaxies \citep{sz1978, FaberGallagher1979, White1991cdm, Kauffmann1993, Johnston1998, Springel2006}. Therefore, a fundamental prediction of this hierarchical paradigm is that satellite galaxies around any massive host present themselves as intact dwarfs, phase-mixed substructures, or in an intermediary stage as stellar streams \citep{BullockJohnston2005, Cooper2010, Cooper2013, Pillepich2015halos, Morinaga2019_MWhalos}. 

Because the evolution of these dwarf galaxy streams and substructures was interrupted due to the shutdown of their star formation at the moment of their accretion, they provide a unique local window into the properties of galaxies that existed at high redshift \citep{Boylan-Kolchin2015nearfield, Boylan-Kolchin2016nearfield}. Also, as these systems sample a wide range of stellar masses ($10^6 \lesssim M_\star/M_\odot < 10^9$; see \citealt{Naidu2022mzr} and \citealt{Sharpe2022halo}), they provide a laboratory for us to test the universality of galaxy formation/evolution processes as we can compare them with observations of intact dwarfs in the Local Group (\citealt{Tolstoy2009} and \citealt{simon2019} for reviews). Conveniently, this approach comes with the advantage that member stars of these disrupted dwarfs are typically much closer and, hence, brighter than their counterparts located in 
surviving Milky Way satellites. This makes them more easily accessible to high-resolution ($\mathcal{R} \geq 20{,}000$) spectroscopy, from which detailed chemical abundances can be obtained for significant amounts of stars, allowing us to understand the properties of their progenitor systems. This ``near-field'' approach to galaxy evolution at the smallest scales is even more appealing given that not even current and future facilities (such as \textit{JWST} or 20--30\,m ground telescopes) will be able to spatially resolve such low-mass galaxies at the redshift range ($0.5 < z \leq 2.0$) probed by these halo debris \citep[][see also \citealt{Naidu2021simulations} for a relevant discussion]{myeongStreamsAndClumps, myeongSequoia, koppelmanHelmi, Forbes2020, Kruijssen2020kraken, naidu2020, Callingham2022gcs}.

The advent of astrometric information for more than a billion stars due to the \textit{Gaia} space mission \citep{GaiaMission, gaiadr1, gaiadr2, GaiaEDR3Summary, GaiaDR32022arXiv}, in particular its second and third data releases (DR2 and DR3, respectively), in combination with large-scale photometric and spectroscopic surveys, has allowed the discovery of a myriad of accreted substructures in the Galactic halo (see \citealt{koppelman2019}, \citealt{Malhan2022atlas} and the stellar-stream compilation by \citealt{Mateu2023galstreams}). Out of these, the most well-characterized disrupted dwarfs with detailed chemistry are, by far, \textit{Gaia}-Sausage/Enceladus \citep[GSE;][]{belokurov2018, Haywood2018, helmi2018} and Sagittarius stream \citep[e.g.,][]{Majewski2003}, the tidal tails of Sagittarius dwarf spheroidal (dSph) galaxy \citep{Ibata1994}. Examples of such efforts include, but are not limited to, \citet{Monty2020}, \citet{Aguado2021SausageSequoia}, \citet{Matsuno2021gseRprocess}, and \citet{Buder2022halo} for GSE, \citet{Hasselquist2019sgr} and \citet{Hayes2020} for Sagittarius, and \citet{Hasselquist2021dwarf_gals} and \citet{Horta2023haloSubs} for both. Apart from these major substructures, other disrupted dwarfs with available high-resolution spectroscopy include Helmi streams (\citealt{helmi1999}; see \citealt{Roederer2010}, \citealt{Aguado2021}, \citealt{Limberg2021hstr}, and \citealt{Matsuno2022hstr} for abundances), Orphan stream (\citealt{Belokurov2006Streams, Belokurov2007Orphan}; e.g., \citealt{Casey2014orphan} and \citealt{Hawkins2023orphan}), and Sequoia\footnote{Although we list Sequoia as an independent disrupted dwarf, we recognize the current dispute in the literature regarding whether or not this population could simply be part of the more massive GSE \citep{Koppelman2020MassiveMerger, Amarante2022gsehalos, Limberg2022gse, Horta2023haloSubs}.} (\citealt{myeongSequoia}; \citealt{matsuno2019, Matsuno2022seq}). For smaller stellar streams, the largest homogeneous study was presented by \citet{Ji2020streams}.

In this contribution, we present the first detailed chemical abundance analysis of Wukong/LMS-1 \citep{naidu2020, Yuan2020lms1, Malhan2021lms1}. This substructure was identified by \citet{Yuan2020lms1} who named it the ``low-mass stellar-debris stream 1'' (LMS-1). Independently, \citet{naidu2020} identified ``Wukong''\footnote{Named after Sun Wukong, the celestial Monkey King from \textit{Journey to the West}. See \citet{naidu2020} for the complete rationale.}, a prominent group of stars in integrals-of-motion space apparently dissociated from any previously known disrupted dwarf. The connection between LMS-1 and Wukong was quickly recognized for their indistinguishable dynamics, including the association with at least two globular clusters, NGC 5024 (M53) and NGC 5053, and two other stellar streams, Indus and Jhelum \citep{Shipp2018, Bonaca2019indus+jhelum, Bonaca2021streams, Malhan2022atlas}. Out of the known disrupted dwarf galaxies, some listed above, Wukong/LMS-1 is especially interesting due to its predicted relatively high stellar mass of ${\sim}10^7\,M_\odot$ \citep{Malhan2021lms1}, which is similar to classical Milky Way satellites such as Sculptor and Fornax dSph galaxies \citep[][]{McConnachie2012catalog}. Therefore, our goal is to constrain Wukong/LMS-1's star-formation history, its production of the heaviest elements via prompt and/or delayed sources of neutron-capture processes, and even look for signatures of dissolved globular clusters in it. Then, we put our results in context by comparing with known Milky Way satellite galaxies of similar mass as well as chemical-evolution models.




This work is organized as follows. Section \ref{sec:data} includes all things related to our observations, data reduction, and radial velocity (RV) measurements. Our methodology for obtaining stellar parameters and abundances from high-resolution spectra is described in Section \ref{sec:methods}. Section \ref{sec:results} is reserved for the presentation of our results. In Section \ref{sec:conclusions}, we provide our concluding remarks and a brief discussion. 

\begin{table*}
    \footnotesize
    \centering
    \setlength{\tabcolsep}{3.0pt}
    \begin{tabular}{cc|cc|c|cc|cc|cc|cc|cc}
    \hline 
Star & \textit{Gaia} DR3 \code{source\_id} & R.A. & Decl. & RV$_{\rm MIKE}$ & $S/N$ & $S/N$ & $\Teff$ & $\sigma_{\Teff}$ & $\logg$& $\sigma_{\logg}$ & $\vt$ & $\sigma_{\vt}$ & [M/H] & $\sigma_{\mbox{[M/H]}}$ \\
& & (deg) & (deg) & (km\,s$^{-1}$) & (4500{\AA}) & (6500{\AA}) & (K) & (K) & (cgs) & (cgs) & (km\,s$^{-1}$) & (km\,s$^{-1}$) & & \\ \hline
Wuk\_1$\phantom{0}$   & 2721020906259820416 & 332.6634 & $\phantom{-}$6.6813$\phantom{0}$   & $-$245.22 & 44 & 85 & 4887 & 103 & 2.12 & 0.50  & 1.58 & 0.11 & $-$1.75 & 0.24 \\
Wuk\_2$\phantom{0}$   & 2671790891601429376 & 322.1504 & $-$5.4919$\phantom{0}$  & $-$96.59$\phantom{0}$  & 30 & 68 & 4623 & 102 & 1.48 & 0.50  & 2.14 & 0.12 & $-$1.91 & 0.25 \\
Wuk\_3$\phantom{0}$   & 2616761777740644864 & 329.9007 & $-$10.9341 & $-$173.24 & 31 & 60 & 4866 & 104 & 1.90  & 0.51 & 1.62 & 0.12 & $-$2.32 & 0.26 \\
Wuk\_4$\phantom{0}$   & 4424978984005490304 & 240.3206 & $\phantom{-}$3.6775$\phantom{0}$   & $\phantom{-}$87.79$\phantom{0}$   & 30 & 60 & 5101 & 118 & 2.30  & 0.53 & 1.79 & 0.11 & $-$2.89 & 0.24 \\
Wuk\_5$\phantom{0}$   & 6337489231846231296 & 222.7506 & $-$4.8440$\phantom{0}$   & $\phantom{-}$2.70$\phantom{00}$     & 37 & 72 & 4670 & 102 & 1.50  & 0.50 & 1.59 & 0.11 & $-$2.41 & 0.25 \\
Wuk\_6$\phantom{0}$   & 4415232603696493696 & 230.8720 & $-$2.0148$\phantom{0}$  & $-$180.10  & 30 & 59 & 5307 & 148 & 3.15 & 0.57 & 1.51 & 0.16 & $-$3.57 & 0.22 \\
Wuk\_7$\phantom{0}$   & 3614246079443061376 & 210.9729 & $-$11.2984 & $-$56.56$\phantom{0}$  & 21 & 46 & 5095 & 159 & 2.43 & 0.59 & 0.93 & 0.10  & $-$2.93 & 0.28 \\
Wuk\_8$\phantom{0}$   & 3635197617107624448 & 201.3197 & $-$4.8245$\phantom{0}$  & $\phantom{-}$233.76  & 34 & 56 & 5312 & 111 & 2.75 & 0.51 & 1.43 & 0.12 & $-$1.98 & 0.26 \\
Wuk\_9$\phantom{0}$   & 3625337025031068544 & 199.5284 & $-$7.8565$\phantom{0}$  & $\phantom{-}$98.30$\phantom{0}$    & 30 & 53 & 4974 & 106 & 2.22 & 0.51 & 1.80  & 0.11 & $-$2.01 & 0.26 \\
Wuk\_10 & 3810150429850681984 & 168.4136 & $\phantom{-}$0.8518$\phantom{0}$   & $\phantom{-}$193.73  & 39 & 64 & 4814 & 101 & 1.90  & 0.50 & 1.55 & 0.11 & $-$1.98 & 0.23 \\
Wuk\_11 & 1225051430189620864 & 213.3937 & $\phantom{-}$10.0073  & $\phantom{-}$4.00$\phantom{00}$       & 39 & 71 & 4706 & 102  & 1.18 & 0.50  & 1.96 & 0.11 & $-$2.37 & 0.23 \\
Wuk\_12 & 3727823702151504896 & 209.6407 & $\phantom{-}$12.8460   & $\phantom{-}$25.68$\phantom{0}$   & 33 & 58 & 4842 & 103 & 1.82 & 0.50  & 1.77 & 0.11 & $-$2.09 & 0.25 \\
Wuk\_13 & 3737533184394415232 & 197.0257 & $\phantom{-}$12.7782  & $\phantom{-}$93.73$\phantom{0}$   & 34 & 61 & 4923 & 103 & 2.15 & 0.50  & 1.70  & 0.11 & $-$1.87 & 0.27 \\
Wuk\_14 & 3696527104395430016 & 187.9680 & $\phantom{-}$0.0382$\phantom{0}$   & $\phantom{-}$49.87$\phantom{0}$   & 43 & 68 & 5099 & 100 & 2.60  & 0.51 & 1.60  & 0.14 & $-$1.23 & 0.22 \\ \hline
    \end{tabular}
    \caption{Observational information, RVs, and stellar parameters for Wukong/LMS-1 stars analysed in this work. The reported $S/N$ values are per pixel. Model atmosphere metallicity values ([M/H]) can be larger than [Fe/H] by up to 0.03\,dex.}
    \label{tab:obs_mike}
\vspace{-3mm}
\end{table*}

\section{Data} \label{sec:data}

\begin{figure}
\centering
\includegraphics[width=1.0\columnwidth]
{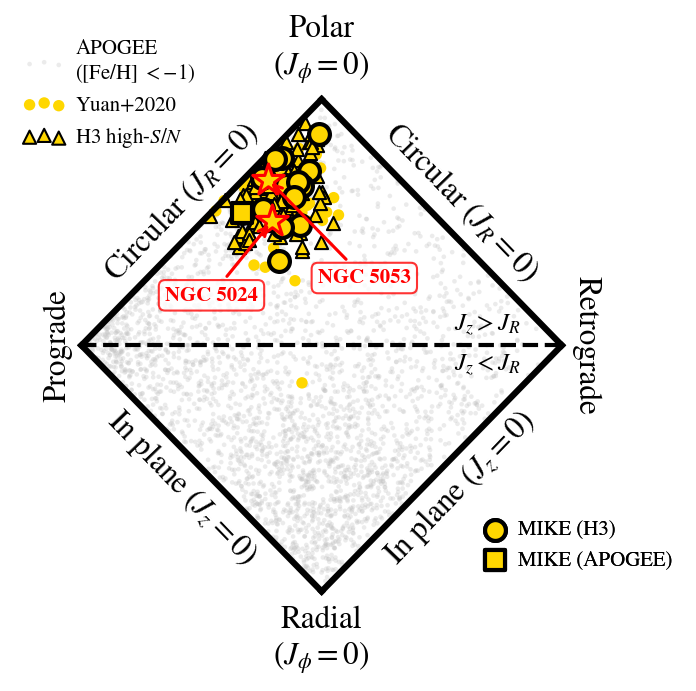}
\caption{Wukong/LMS-1 stars in projected action-space diagram. The horizontal axis corresponds to $J_\phi/J_{\rm total}$, where $J_{\rm total} = J_R + |J_\phi| + J_z$ (see text). The vertical axis is $(J_z - J_R)/J_{\rm total}$. Larger symbols with thick black edges are those observed with MIKE; circles for H3 targets (Wuk\_1 through \_13) and square for the APOGEE one (Wuk\_14). Triangles with thin black edges are the ``high-$S/N$'' sample of Wukong/LMS-1 members from \citet{Johnson2022wukong}. Smaller circles without edges are the ``LMS-1'' stars from \citet[][see text]{Yuan2020lms1}. Grey dots are metal-poor ($\rm[Fe/H] < -1$) stars from APOGEE DR17. Star symbols with red edges are the Wukong/LMS-1 globular clusters NGC 5024 (M53) and NGC 5053.
\label{fig:actions}}
\end{figure}

\subsection{Observations} \label{sec:obs}
\indent\indent We observed 13 Wukong/LMS-1 targets (Wuk\_1 to \_13)\footnote{Throughout this work, we use the ``Wuk'' prefix followed by a natural number as unique names for stars in our sample (see Table \ref{tab:obs_mike}).}, which were selected as best-suited for covering a wide metallicity range within the telescope time allocation available. These were originally identified by \citet{naidu2020} using data from the Hectochelle in the Halo at High Resolution (H3) survey \citep[][]{Conroy2019surveyH3}, including orbital energy and angular momentum criteria, but also an $\rm[Fe/H] < -1.45$ cut to avoid contamination by GSE stars. The observed sample is also contained in the ``high signal-to-noise'' (high-$S/N$) sample of Wukong/LMS-1 stars from \citet{Johnson2022wukong}, also with H3 data, who refined \citeauthor{naidu2020}'s (\citeyear{naidu2020}) selection. All stars are also bright enough ($15 <$ \textit{Gaia}'s $G$ $\lesssim 16$) for high-resolution spectroscopy with our setup (see below). We collected spectra for an extra metal-rich ($\rm[Fe/H] \sim -1.3$) Wukong/LMS-1 star (Wuk\_14; Table \ref{tab:obs_mike}) at $G \approx 15$ found in the Apache Point Observatory Galactic Evolution Experiment \citep[APOGEE;][]{apogee2017} DR17 catalog \citep[][]{APOGEEdr17}. In order to guarantee that this additional target is a genuine Wukong/LMS-1 member, not a GSE interloper, we searched APOGEE for stars that respected the combined criteria from \citet{naidu2020} and \citet{Johnson2022wukong} as well as \citet{Yuan2020lms1}, which was developed independently from the H3 survey papers. The full criteria is written below. 
Because GSE-like mergers are not expected to deposit debris on such acute polar orbits \citep[see][\citealt{Limberg2023sgr} for discussion]{Amarante2022gsehalos}, we can be confident that our APOGEE metal-rich target is a genuine member of Wukong/LMS-1. 

In Figure \ref{fig:actions}, we exhibit all our targets in projected action space within the \citet{mcmillan2017} Galactic model potential. The action vector is written as $\mathbf{J} = (J_R, J_\phi, J_z)$, where $J_R$, $J_\phi$, and $J_z$ are the radial, azimuthal, and vertical components in a cylindrical frame. Both of the above-mentioned Wukong/LMS-1 samples and the globular clusters NGC 5024 and NGC 5053 are also plotted. For these orbit calculations, we integrated for 20\,Gyr forward using the \code{AGAMA} library \citep{agama}. Positions and proper motions on the sky are from \textit{Gaia} DR3. RVs were determined by us for the stars we observed (see below). Other RVs employed were taken from their parent samples. Distances come from spectro-photometric fits, \code{MINESweeper} \citep{Cargile2020minesweeper} for H3 and \code{StarHorse} \citep{Queiroz2020, Queiroz2023starhorse} for APOGEE. For NGC 5024 and NGC 5053, all phase-space information is from \citet{VasilievBaumgardt2021gcs}. We realized each orbit 100 times in a Monte Carlo scheme assuming Gaussian uncertainties for these quantities. The final adopted values are the medians of the resulting distributions. The adopted distance from the Sun to the Galactic center is 8.2\,kpc \citep[][]{BlandHawthorn2016}, the circular velocity at this position is 232.8\kms \citep{mcmillan2017}, and the assumed peculiar motion of the Sun is $(U,V,W)_\odot = (11.10, 12.24, 7.25)$\kms \citep{schon2010}.

We note that the model potential as well as Galactic fundamental parameters adopted here are different from previous H3 survey works \citep{naidu2020, Johnson2022wukong}. Therefore, the criteria used by these authors to select Wukong/LMS-1 stars become slightly different after our above-described calculations. In the spirit of making this paper self-sufficient, we provide updated values for the kinematic/dynamical quantities that define the Wukong/LMS-1 structure below. Nevertheless, we reinforce that the original target selection was made simply based on the H3 survey samples from \citet{naidu2020} and \citet{Johnson2022wukong} plus covering the largest possible metallicity range within the available telescope allocation.

Following \citet{Johnson2022wukong}, we have:
\begin{itemize}
    \item $(J_z - J_R)/J_{\rm total} > 0.3$\footnote{We call the attention to the difference in definition of $J_{\rm total}$ between ours and \citeauthor{naidu2020}'s (\citeyear{naidu2020}) work, also \citet{Johnson2022wukong}. These authors assumed the vectorial definition $J_{\rm total} = \sqrt{J_R^2 + J_\phi^2 +  J_z^2}$.} and
    \vspace{+1mm}
    \item $90^\circ < \theta < 120^\circ$,
\end{itemize}
where $J_{\rm total} = J_R + |J_\phi| + J_z$ (Figure \ref{fig:actions}) and $\theta = \arccos{(L_z/L)}$ refers to the orbital inclination, which characterizes the direction of the angular momentum vector $\mathbf{L} = (L_x, L_y, L_z)$ in a Galactic Cartesian frame. Within these definitions, $L = \sqrt{L_x^2 + L_y^2 + L_z^2}$ is the total angular momentum and $L_z \equiv J_{\phi}$ is the vertical component of it. These criteria are accompanied by those from \citet{naidu2020}:
\begin{itemize}
    \item $-1000 < L_z/({\rm kpc \kms}) < 0$ and
    
    \vspace{+1mm}
    \item $E < -1.15 \times 10^5\,{\rm km^2\,s^{-2}}$,
\end{itemize}
where $E$ is the total orbital energy. Apart from these cuts, \citet{naidu2020} also removed possible Sagittarius stream interlopers using the simple method of \citet{Johnson2020sgr}, which has been shown to be likely complete \citep{Penarrubia2021sgr}. Hence, when selecting \textit{for} Wukong/LMS-1, one can use
\begin{itemize}
    \item $L_y > -2000\,{\rm kpc \kms}$,
\end{itemize}
fully eliminating Sagittarius contamination within its $L_z$ range. \citet{naidu2020} also included a cut in orbital eccentricity to eliminate GSE stars. However, this is redundant with the action-space selection of \citet{Johnson2022wukong} that requires $J_z > J_R$.

\begin{figure}
\centering
\includegraphics[width=1.0\columnwidth]{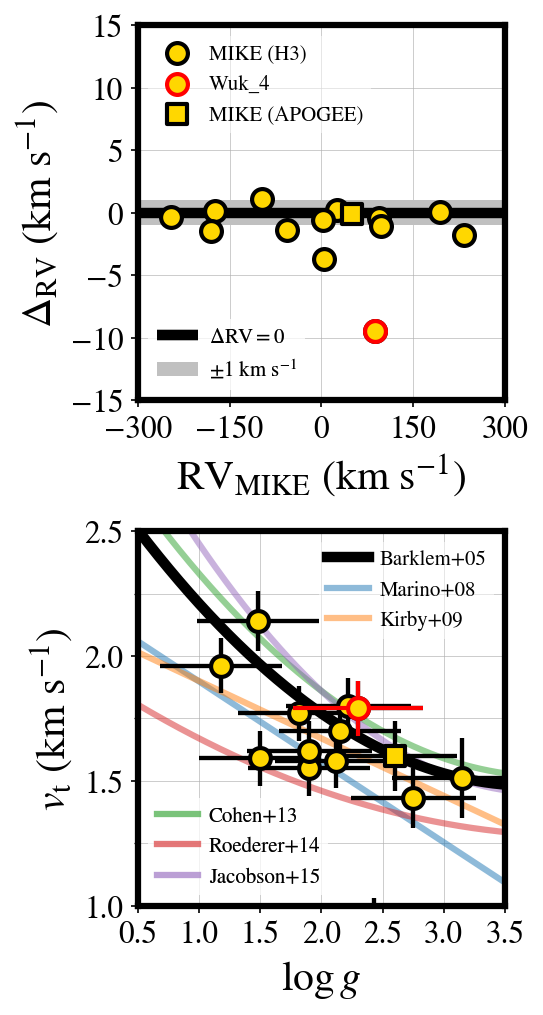}
\caption{Top panel: comparison between MIKE and survey RVs, either H3 (circles) or APOGEE (square). The vertical axis shows $\Delta_{\rm RV}$, which corresponds to MIKE RVs minus H3/APOGEE ones. 
The black line shows $\Delta_{\rm RV} = 0$ and the grey shaded region covers ${\pm}1$\kms, the systematic RV uncertainty for MIKE measurements. Bottom panel: \logg versus \vt relation (see text). Colored lines are empirical quadratic fits to various samples of low-metallicity stars \citep[][see \citealt{AlexJi2023ret2}]{barklem2005, Marino2008m4, Kirby2009sculptor, Cohen2013, roederer2014, Jacobson2015metalpoor}. The black line was used to determine the \vt for the most metal-poor star analysed (Wuk\_6; see text). Wuk\_4 is shown with red edges in both panels (see Section \ref{sec:cemp}).
\label{fig:rv}}
\vspace{-2mm}
\end{figure}

We observed all the Wukong/LMS-1 stars with the Magellan Inamori Kyocera Echelle \citep[MIKE;][]{Bernstein2003mike} spectrograph installed in the Magellan Clay telescope (6.5\,m) located at Las Campanas Observatory, Chile. For all stars over all observing runs (June 2022 for H3 follow-up and May 2023 for APOGEE), we used $0.7\arcsec$ slit and $2{\times}2$ on-chip binning. This configuration leads to resolving powers of $\mathcal{R} \sim 35{,}000$ and $28{,}000$ for the blue (wavelength $\lambda < 5000$\,\AA) and red ($\lambda > 5000$\,\AA) arms of MIKE spectra, respectively. All data were reduced using 
the \texttt{CarPy}\footnote{\url{https://code.obs.carnegiescience.edu/mike}.} package \citep{carpy}. The final $S/N$ reached was typically 30--40 per pixel at 4500\,\AA \ and 50--70 at 6500\,\AA \ (Table \ref{tab:obs_mike}). 

\subsection{Radial velocities and spectra normalization} \label{sec:rv}
\indent\indent We derived RVs for all Wukong/LMS-1 stars by cross-correlating against a high-$S/N$ MIKE spectrum of the metal-poor standard HD 122563 using the Labeling Echelle Spectra with SMHR and Payne (\texttt{LESSPayne}\footnote{\url{https://github.com/alexji/LESSPayne}.}; A. P. Ji, in preparation) code. In a nutshell, \texttt{LESSPayne} combines Spectroscopy Made Harder (\texttt{smhr}\footnote{\url{https://github.com/andycasey/smhr}.}; \citealt{Casey2014smhr}) with \texttt{Payne4MIKE}\footnote{\url{https://github.com/alexji/Payne4MIKE}.} \citep{Ting2019ThePayne}\footnote{\url{https://github.com/tingyuansen/Payne4MIKE}} and consolidates it into a single package. As of now, the RV measurement routine within \code{LESSPayne} is a carbon-copy of \code{smhr}'s. The formal statistical uncertainty of RVs from MIKE spectra (``${\rm RV}_{\rm MIKE}$'' in the top panel of Figure \ref{fig:rv} and Table \ref{tab:obs_mike}) could, in principle, reach ${\sim}0.1$\kms. However, a systematic error of ${\sim}1$\kms is introduced due to slit centering and wavelength calibration \citep[see discussion by][]{Ji2020MagLiteS}. For the purpose of this work, we are satisfied that the membership of our stars does not depend on the choice of RV value, either MIKE or H3/APOGEE. Indeed, at this RV precision, distances are the dominant source of errors for orbital parameters
. The maximum difference between our MIKE RVs and H3 ones is ${\approx}10$\kms (Wuk\_4), but is usually ${<}2$\kms (Figure \ref{fig:rv}). See Section \ref{sec:cemp} for the possibility that Wuk\_4 is in a binary system. For the APOGEE star Wuk\_14, the difference in RV is 0.5\kms.

With the reduced data and RVs at hand, we proceed to stitch orders and normalize the MIKE spectra using cubic spline functions. 
\lp initializes \code{smhr}-like files from the best-fitting \code{Payne4MIKE} synthetic spectrum. This procedure drastically accelerates the normalization and equivalent width measurements, as it identifies where absorption features occur and masks them. This method is identical to the one described in \citet{AlexJi2023typhon}, with the difference that, now, the whole process has been packaged into \lp. The entire spectrum was inspected for all stars, but the \lp continuum needed minor fixes only at the bluest orders, where $S/N$ is lower, or when prominent wide absorption was present, such as \ion{Ca}{II} K/H (3900--4000\,\AA) and/or the C-H \textit{G} band (${\sim}4300$\,\AA). 

\begin{figure*}
\centering
\includegraphics[width=2.0\columnwidth]{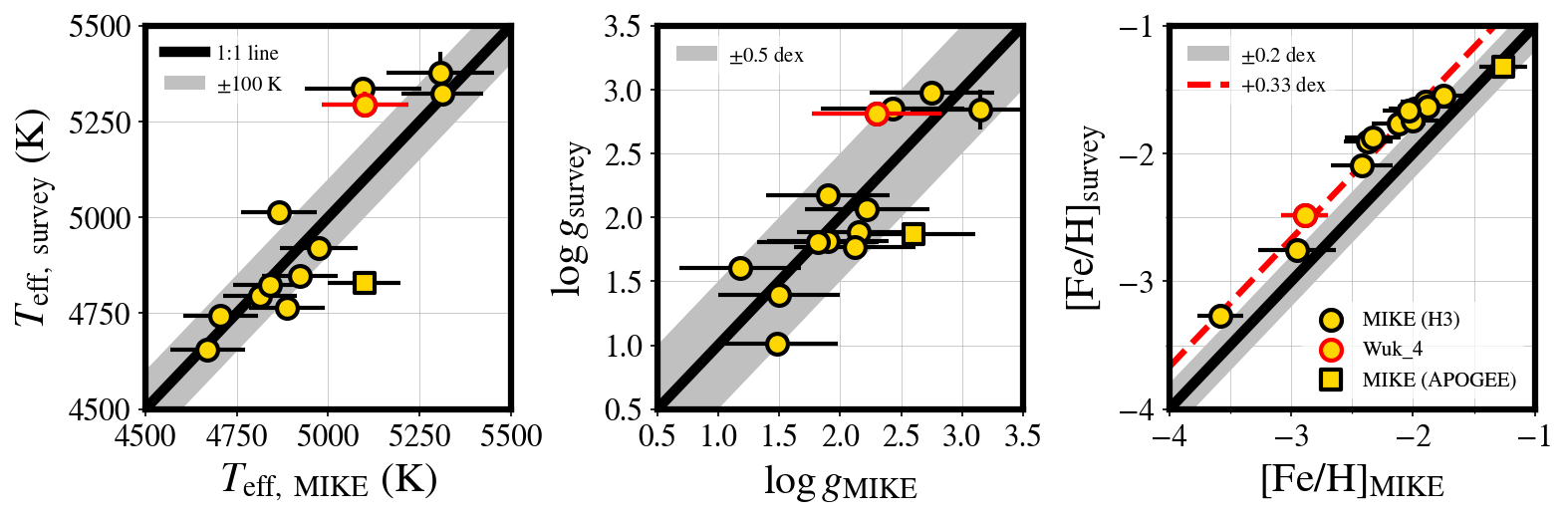}
\caption{Comparisons between our derived stellar parameters from MIKE spectra (horizontal axes; Table \ref{tab:obs_mike}) and the values from the spectroscopic surveys (vertical), either H3 (circles) or APOGEE (square). Left: \Teff. Middle: \logg. Right: [Fe/H]. One-to-one lines are shown in black. Grey shaded areas represent the systematic uncertainties adopted. The red line in the right panel shows the average difference between our derived metallicities and those from H3.
\label{fig:comp_h3}}
\end{figure*}

\section{Methods} \label{sec:methods}
\indent \indent For this project, we considered the atomic and molecular line list from \citet{Ji2020streams}, which was adapted from the \code{linemake}\footnote{\url{https://github.com/vmplacco/linemake}.} compilation \citep{Placco2021linemake}. This line list was also used by \citet{Ji2020streams} to analyze the Indus and Jhelum streams that could be associated with Wukong/LMS-1 and which we compare to in Section \ref{sec:results}. We rejected all lines at ${<}3860$\,\AA \ in our analysis due to the lower $S/N$ at the bluest portions of the MIKE spectra
. For our equivalent width measurements, we assumed Gaussian profiles. Lines that required Voigt profiles were rejected, with the exception of those in the \ion{Mg}{I} triplet (5150--5200\,\AA). Abundances for Mg from these strong lines are consistent with others considered. Overall, we fitted between 70 (for our most metal-poor star) and 380 (most metal-rich) lines in each MIKE spectrum. Individual line measurements are provided in supplementary online tables.

\subsection{Stellar Parameters}
\indent\indent We computed fully spectroscopic stellar parameters for all stars under the assumption of local thermodynamic equilibrium (LTE) using \lp, which wraps the radiative transfer code \code{MOOG} \citep{Sneden1973moog}, specifically a well-tested version\footnote{\url{https://github.com/alexji/moog17scat}.} that includes appropriate treatment of scattering \citep{Sobeck2011}. We employed $\alpha$-enhanced ($\rm[\alpha/Fe] = +0.4$) 1D plane-parallel model atmospheres \citep{Castelli2003atmospheres} in our analysis. 

Effective temperatures (\Teff) were estimated by balancing the abundances of \ion{Fe}{I} lines against their excitation potential. Surface gravity values (\logg) were found by minimizing the difference between \ion{Fe}{I} and \ion{Fe}{II} abundances. Microturbulence velocities (\vt) were determined by reducing the trend between \ion{Fe}{II} abundances and reduced equivalent widths; \ion{Fe}{II} lines are preferred as they cover a wider range of reduced equivalent width values in red giants and are less affected by non-LTE (NLTE) effects \citep[see][]{Ji2020streams}. For the most metal-poor star in our sample (Wuk\_6; model atmosphere metallicity $\rm[M/H]=-3.57$), only three \ion{Fe}{II} lines were available. Hence, for this star, we obtained \vt from the empirical relation with \logg based on the work of \citet[][see appendix B of \citealt{AlexJi2023ret2} and bottom panel of Figure \ref{fig:rv}]{barklem2005}.

Lastly, we recalibrated the \Teff for each star to the ``photometric scale'' of \citet{Frebel2013photscale} and, then, rederived \logg and \vt by repeating the same steps described above. We reanalysed from scratch a couple of stars from \citet{Ji2020streams} with our method and verified that these authors' photometric stellar parameters are compatible ($1\sigma$) with those revised with the \citet{Frebel2013photscale} correction. On average, \Teff values become $\approx$200\,K hotter after this step. For \logg and [M/H], the corrected values are, on average, 0.67\,dex and 0.24\,dex larger, respectively. For \vt, the final values are 0.07\kms lower on average. Note that we do \textit{not} just apply constant offsets, but rather the stellar parameters of each star change independently according to their recalibrated \Teff and these listed values are simply the average corrections. In Table \ref{tab:obs_mike} (also all figures), we provide only corrected parameters. Stellar parameters prior to recalibration are provided as online supplementary material. Throughout the remainder of this work, we consider only recalibrated stellar parameters, including for the abundance analysis (Section \ref{sec:abund}).

Apart from just statistical uncertainties, which come from the fitting process of individual lines as well as from the slopes of the excitation/ionization balance fits, it is crucial to consider systematics that can affect our stellar parameters and which are propagated into abundances. Systematic errors in stellar spectroscopic analysis come, in principle, from departures from the 1D LTE assumption, i.e., 3D and/or NLTE effects, which are especially relevant for metal-poor giants \citep{Asplund2005nlte}. In order to obtain empirically motivated systematic errors, we looked at the sample of low-metallicity stars analysed by \citet{ezzedine2020}, which spans a wide range of stellar parameters that completely encompasses our sample; $4600 < \Teff/{\rm K} \lesssim 5300$, $1 < \logg \lesssim 3$, $-3.50 \lesssim {\rm[Fe/H]} < -1.25$, and $1 < v_{\rm t}/({\rm km \ s^{-1}}) \lesssim 2$. These authors provided different sets of stellar parameters for each of their stars, including 1D LTE with the \citet{Frebel2013photscale} correction and 1D NLTE. We verified that the average differences between these stellar parameter estimates are 100\,K, 0.5\,dex, 0.1\kms, and 0.2\,dex for \Teff, \logg, \vt, and [Fe/H] and these values are adopted as systematic errors, summing in quadrature with the statistical ones to obtain final uncertainties for these quantities. For Wuk\_6, we adopted a \vt statistical uncertainty of 0.12\kms, which is the intrinsic scatter from the quadratic fit to the \logg versus \vt relation
. We note that, despite the apparently excessively large \logg errors, \Teff continues to have the most important impact on abundance uncertainties.

A comparison between our MIKE parameters and those from the H3 survey/\code{MINESweeper} is provided in Figure \ref{fig:comp_h3}. For \Teff and \logg, these are compatible ($1\sigma$) for all stars. For [Fe/H], H3 values are larger by ${\sim}0.3$\,dex, but this level of systematics can be attributed to NLTE effects \citep[see figure 2 of][]{ezzedine2020}. In comparison to APOGEE parameters, we found \Teff and \logg differences of ${\sim}300$\,K and ${\sim}0.7$\,dex, respectively, but with equivalent [Fe/H] values (also Figure \ref{fig:comp_h3}). We note, however, that, before applying the \citet{Frebel2013photscale} photometric correction, our purely spectroscopic \Teff and \logg would be compatible ($1\sigma$) with APOGEE values. With respect to the H3 values, this survey adopts spectro-photometric fits to obtain stellar parameters from broad-band photometry and model isochrones. Henceforth, it is not unexpected that our photometrically recalibrated \Teff and \logg are similar, as is the case between ours and \citeauthor{Ji2020streams}'s (\citeyear{Ji2020streams}) photometric stellar parameters.

\begin{figure*}
\centering
\includegraphics[width=1.98\columnwidth]{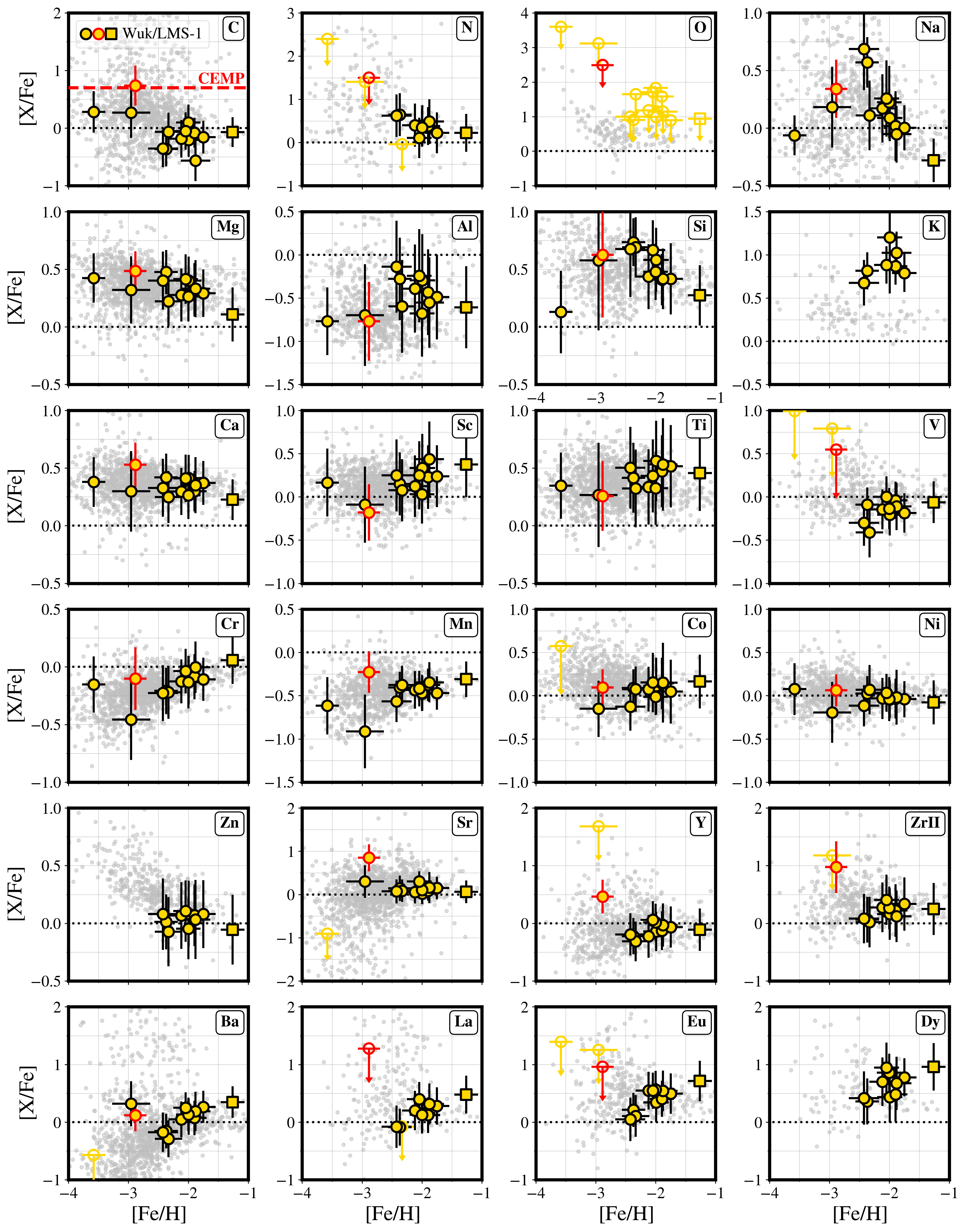}
\caption{[X/Fe] versus [Fe/H] plots for all elements estimated in this work for our Wukong/LMS-1. In this figure, Fe is always \ion{Fe}{I}. The top right corner of each panel shows the X element represented. Filled symbols are measured abundances, while open ones are upper limits. Circles correspond to stars followed-up from H3 (Wuk\_1 to \_13) and the square is from APOGEE (Wuk\_14). Dotted lines at $\rm[X/Fe] = 0$ show the solar level. The red dashed line at $\rm[C/Fe]=+0.7$ in the top left panel delineates the boundary for the definition of CEMP star (see text). The symbol with red edge is Wuk\_4. In all panels, the grey dots in the background are metal-poor stars from our SAGA data base compilation (see text).
\label{fig:all_abunds}}
\end{figure*}

\subsection{Abundances} \label{sec:abund}

\indent\indent We obtained abundances for up to 29 species of 24 elements. Equivalent widths were used for the abundances of \ion{Na}{I}, \ion{Mg}{I}, \ion{Si}{I}, \ion{K}{I}, \ion{Ca}{I}, \ion{Ti}{I}, \ion{Ti}{II}, \ion{Cr}{I}, \ion{Cr}{II}, \ion{Ni}{I}, \ion{Zn}{I}, and \ion{Sr}{I}, as well \ion{Fe}{I} and \ion{Fe}{II}. For heavily blended features, molecular features, or those requiring hyperfine splitting, we defaulted to spectral synthesis. For the task, \code{LESSPayne} performs a $\chi^2$ minimization over local continuum with RV and abundance as free parameters, as well as a smoothing parameter to account for resolution. This approach was employed for C-H and C-N molecules, \ion{Al}{I}, \ion{Sc}{II}, \ion{V}{I}, \ion{V}{II}, \ion{Mn}{I}, \ion{Co}{I}, \ion{Sr}{II}, \ion{Y}{II}, \ion{Zr}{II}, \ion{Ba}{II}, \ion{La}{II}, \ion{Eu}{II}, and \ion{Dy}{II}. We also synthesized the \ion{Si}{I} line at 3905\AA \ due to substantial blending. Additionally, we applied this technique to derive $5\sigma$ upper limits for \ion{O}{I}, as well as other elements, for all our stars. For C-H and C-N, we adopted $^{12}{\rm C}/^{13}{\rm C} = 9$. For Ba and Eu, solar $r$-process isotopic ratios were assumed \citep{Sneden2008}.  For [Ti/Fe], [V/Fe], [Cr/Fe], and [Sr/Fe] ratios, we adopt \ion{Ti}{II}, \ion{V}{I}, \ion{Cr}{I}, and \ion{Sr}{II} as our fiducial values (Figure \ref{fig:all_abunds}).

Final abundances were calculated as inverse-variance weighted averages, taking into account uncertainties from individual lines for each species. Our adopted procedure is similar to \citet[][]{Ji2020MagLiteS}, which neglects correlations between stellar parameters \citep[][]{McWilliam2013params}. The one difference between our methodology and these authors' is the inclusion of an error floor on a line-by-line basis ($\sigma_{\rm sys}$), which should encapsulate additional unknowns coming from, for instance, atomic data or the 1D model atmospheres \citep[see][]{Ji2020streams}. For this purpose, we adopted a constant $\sigma_{\rm sys} = 0.1$\,dex floor for most lines. For some species with hyperfine splitting, where the smoothing kernel might be degenerate with abundance, such as Sc, Mn, and Ba, we add an extra 0.1\,dex, i.e., $\sigma_{\rm sys} = 0.2$\,dex for all lines. For the C-N molecular band, which is located in the bluest part of MIKE spectra (3865--3885\,\AA) where the $S/N$ is low and the continuum placement is difficult, we implemented $\sigma_{\rm sys} = 0.3$\,dex. The Al line at 3961\,\AA, which is the only one considered for this element, not only suffers from the same caveats as the C-N, but is also at the wing of a hydrogen line (H$\epsilon$), so we employed $\sigma_{\rm sys} = 0.3$\,dex for it as well. 
We provide the relevant equations below, which are analogous to equations 1 to 5 in \citet{Ji2020MagLiteS}, but incorporating our modification and with updated terminology. 

For a certain line $i$ of a given element/specie X, its total uncertainty ($\sigma_i$) of the associated abundance ($A_i$) can be written as the quadrature sum of the different sources of error, i.e.,
\begin{equation}
\sigma_{i}^2 = \sigma_{i, \rm stat}^2 + \sigma_{i, \rm SP}^2 + \sigma_{i, \rm sys}^2,
\label{eq:sigma}
\end{equation}
where $\sigma_{i, \rm stat}$ is the statistical error, which comes from spectrum noise and the line-fitting procedure, and $\sigma_{\rm SP}$ is the total stellar-parameter error budget;
\begin{equation}
\sum_{\rm SP} \delta_{i, {\rm SP}}^2 = \sigma_{i, {\rm SP}}^2 = \delta_{i,\Teff}^2 + \delta_{i,\logg}^2 + \delta_{i,\vt}^2 + \delta_{i,\rm [M/H]}^2.
\label{eq:sigma_sp}
\end{equation}
In this equation, $\delta_{i,\Teff}$, $\delta_{i,\logg}$, $\delta_{i,\vt}$, and $\delta_{\rm i,[M/H]}$ are abundance offsets associated with uncertainties in \Teff, \logg, \vt, and [M/H], respectively, as listed in Table \ref{tab:obs_mike}. Because these quantities retain their sign, we refer to them as ``$\delta_{i, \cdots}$''. Finally, the inverse-variance weights can be assigned to each line, 
\begin{equation}
w_i = \dfrac{1}{\sigma_i^2},
\label{eq:weight}
\end{equation}
and the final abundance can be computed as
\begin{equation}
A({\rm X}) = \dfrac{\sum^N_i w_i A_i}{\sum^N_i w_i},
\label{eq:abund}
\end{equation}
where $N$ is the total amount of lines available for X. 

Uncertainties in $A({\rm X})$, as well as abundance ratios, are propagated in an identical fashion to \citet[][]{Ji2020MagLiteS}. Again, to make this paper self-sufficient, we reproduce these authors' equations 6 through 10 (see their appendix) below.
\begin{equation}
\sigma^2_{\rm stat} = \dfrac{\sum^N_i w_i [A_i - A({\rm X})]^2}{\sum^N_i w_i} + \dfrac{1}{\sum^N_i w_i},
\label{eq:sigma_stat_AX}
\end{equation}
where this total statistical uncertainty for abundance $A({\rm X})$ includes the weighted standard error across abundances $A_i$ for $N$ different lines $i$ of element/specie X as well as spectrum noise. Note that when $N = 1$, the first term of Equation \ref{eq:sigma_stat_AX} goes to zero and, in that situation, the second term is fully responsible for propagating line-by-line errors (including our $\sigma_{\rm sys}$) to the final $\sigma_{\rm stat}$ through the weights pre-computed by Equation \ref{eq:weight}. For stellar parameters,
\begin{equation}
\delta_{\rm SP} = \dfrac{\sum^N_i w_i \delta_{i, {\rm SP}}}{\sum^N_i w_i}.
\label{eq:delta_sp_AX}
\end{equation}
We recall that covariance between stellar parameters (\Teff, \logg, $v_t$, and [M/H]) is neglected \citep[see][]{McWilliam2013params}. Then, the total error budget for an abundance ratio [X/H] between element/specie X and hydrogen is obtained by summing statistical and stellar-parameter uncertainties in quadrature:
\begin{equation}
\sigma^2_{\rm [X/H]} = \sigma^2_{\rm stat} + \sum_{\rm SP} \delta^2_{\rm SP}.
\label{eq:unc_XH}
\end{equation}
Finally, the uncertainty of an abundance ratio between elements/species X and Y accounts for covariance between X and Y through stellar parameters:
\begin{equation}
\sigma^2_{\rm [X/Y]} = \sigma^2_{\rm X,stat} + \sigma^2_{\rm Y,stat} + \sum_{\rm SP} (\delta_{\rm X,SP} - \delta_{\rm Y,SP})^2.
\label{eq:unc_XFe}
\end{equation}

Our abundance inventory is shown in Figure \ref{fig:all_abunds}. Our Wukong/LMS-1 stars (yellow symbols) are plotted against a compilation of metal-poor stars (grey dots) from the Stellar Abundances for Galactic Archaeology (SAGA) database \citep{SAGA_1, SAGA_4}, including the works of \citet{Fulbright2000}, \citet{barklem2005}, \citet{Cohen2013}, \citet{yong2013full}, \citet{roederer2014}, \citet{Jacobson2015metalpoor}, and \citet{Li2022lamost}. All abundance information, both ours and SAGA's, was normalized to the solar composition of \citet{asplund2009}. Whenever we mention the ratio between an element X and Fe ([X/Fe]), we adopt \ion{Fe}{I} as reference. 
We highlight the red dashed line in the upper left panel of Figure \ref{fig:all_abunds}, which delineates the boundary for ``carbon-enhanced metal-poor'' stars (CEMP; $\rm[C/Fe] > +0.7$ and $\rm[Fe/H] < -1$; \citealt{beers2005}, \citealt{aoki2007}, \citealt{frebel2015}, \citealt{frebel2018}). Dotted lines portray the solar-level abundances ($\rm[X/Fe] = 0$) in all panels.

\begin{figure*}
\centering
\includegraphics[width=2.0\columnwidth]{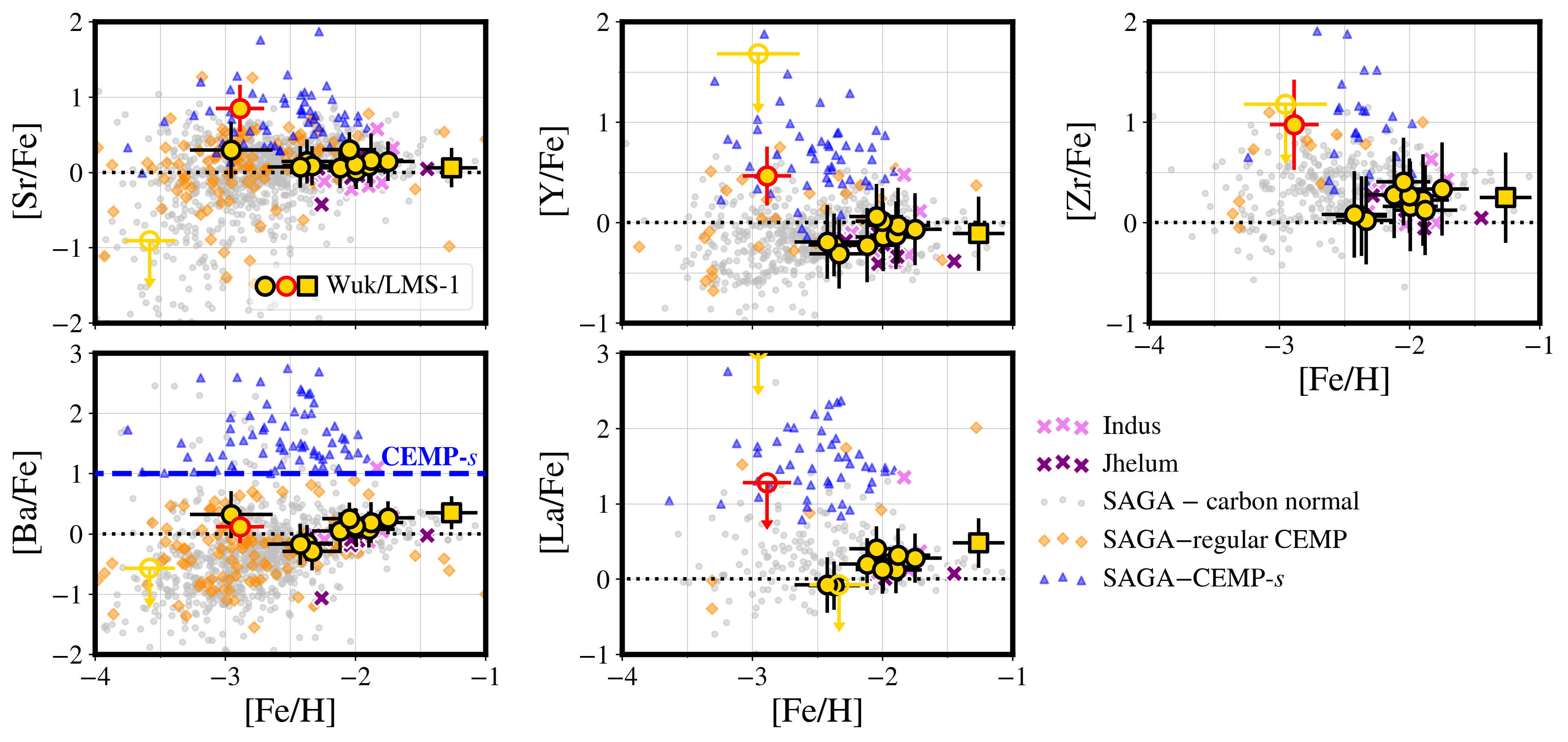}
\caption{Abundance plots for some neutron-capture elements: [Sr, Y, Zr, Ba, La/Fe] versus [Fe/H]. Yellow symbols are our Wukong/LMS-1 stars with detected abundance for the given element in each panel. Empty symbols are upper limits. The marker with red edge represents the CEMP star in our sample (Wuk\_4). The yellow square is the metal-rich star from APOGEE (Wuk\_14). Pink and purple crosses are stars from Indus and Jhelum stellar streams, respectively. Grey dots in the background are the metal-poor stars from the SAGA data base compilation with normal carbon enrichment ($\rm[C/Fe] \leq +0.7$). Regular CEMP stars are marked as orange diamonds ($\rm[C/Fe] > +0.7$ and $\rm[Ba/Fe] \leq +1.0$). CEMP-$s$ stars ($\rm[C/Fe] > +0.7$ and $\rm[Ba/Fe] > +1.0$) are shown in blue triangles (see blue dashed line in the bottom left panel). The solar abundance level is shown as dotted black lines in all panels.
\label{fig:cemp_abunds}}
\end{figure*}

\section{Results} \label{sec:results}

\indent\indent Throughout this section, we compare our derived abundance information for Wukong/LMS-1 with literature results, also with the aid of SAGA data base \citep{SAGA_1, SAGA_4}, but not limited to, for other dwarf galaxies, accreted substructures, and stellar streams. We gathered data for three dSph galaxies, namely Sculptor \citep[Scl,][]{Shetrone2003, Geisler2005, jablonka2015scl, Simon2015scl, Hill2019}, Fornax \citep[Fnx,][]{Shetrone2003, Letarte2010}, and Draco \citep[Dra,][]{Shetrone2001, Cohen2009, Tsujimoto2015}
. We also consider Sagittarius (Sgr) dSph$+$stream \citep[][updated to APOGEE DR17]{Hayes2020}. We also took data for 
Reticulum II \citep[Ret II,][]{ji2016a, Ji2016b} ultra-faint dwarf (UFD). We adopt the sample of ``low-$\alpha$'' stars from \citet{NissenSchuster2010, NissenSchuster2011}, representing GSE, as an example of phase-mixed accreted dwarf.

For Indus and Jhelum stellar streams, we use the line measurements from \citet{Ji2020streams}, but we recomputed the average abundance ratios using our described methodology (Section \ref{sec:abund}) rather than the weights from their paper. We proceed into this section with the assumption that our analysed abundances for Wukong/LMS-1, Indus, and Jhelum are on a consistent scale and with the understanding that these stellar streams are all associated through a common progenitor dwarf galaxy. If these hypotheses hold true, the bulk of their chemical-abundance patterns should be effectively indistinguishable from each other. Having said that, we caution that, although chemistry is a great tool for falsifying associations between substructures, similar abundance patterns do not automatically confirm common origins, especially if these are not too different from the underlying halo population.

\subsection{A CEMP star with peculiar neutron-capture signatures} \label{sec:cemp}

\indent\indent It has been known for at least a couple of decades that the fraction of CEMP stars increase as function of decreasing metallicity \citep[][and see \citealt{Arentsen2022cemp} for a compilation]{beers1992, norris1997, rossi1999, rossi2005, lucatello2006, lee2013, placco2014Carbon, placco2018, yoon2018}. The evolution of the CEMP fraction with [Fe/H] appears to be identical between the Milky Way's halo and UFD galaxies \citep{Ji2020MagLiteS}. However, detailed studies for some more massive dwarfs, such as classical dSph ones, show that some discrepancy might exist with the Milky Way, in particular for Sculptor \citep{Skuladottir2015cempScl, Skuladottir2021UMPsculptor, Skuladottir2023Scl, chiti2018} and, possibly, Sagittarius \citep[][but see \citealt{Limberg2023sgr}]{Chiti2019sgr, Chiti2020sgr}.

In this context, we confirm that Wuk\_4, one of our most metal-poor Wukong/LMS-1 members ($\rm[Fe/H] = -2.89\pm0.19$), is a CEMP star ($\rm[C/Fe] = +0.74\pm0.25$; top left panel of Figure \ref{fig:all_abunds}, symbol with red edge). The carbon-abundance correction for evolutionary effects is only ${+}0.01$\,dex \citep{placco2014Carbon}. After, applying the carbon correction to all other stars in our Wukong/LMS-1 sample, none of them turned out to be CEMP. The corrected [C/Fe] values for all stars are provided alongside the full abundance table as supplementary material. Wuk\_4 had already been identified as a CEMP candidate by \citet{Lucey2023cemp} through spectro-photometric data from \textit{Gaia} DR3. Wuk\_4 also shows strong enhancement in Sr, Y, and Zr ($+0.5 \lesssim \rm[X/Fe] \leq +1.0$; upper panels of Figure \ref{fig:cemp_abunds}). Interestingly, this is not accompanied by significant enrichment in either Ba ($\rm[Ba/Fe] = +0.12\pm0.24$) or La (only upper limit found) as would be expected if Wuk\_4 was a ``typical'' CEMP-$s$ star \citep{beers2005, frebel2018}, i.e., a CEMP star also enhanced in the slow ($s$-) neutron capture process 
(defined as $\rm[Ba/Fe] > +1.0$). Apart from Wuk\_4, other Wukong/LMS-1 stars have effectively identical chemical compositions to the bulk of Indus and Jhelum stars in all abundance panels in Figure \ref{fig:cemp_abunds} (Sr, Y, Zr, Ba, La).

\begin{figure*}
\centering
\includegraphics[width=2.0\columnwidth]{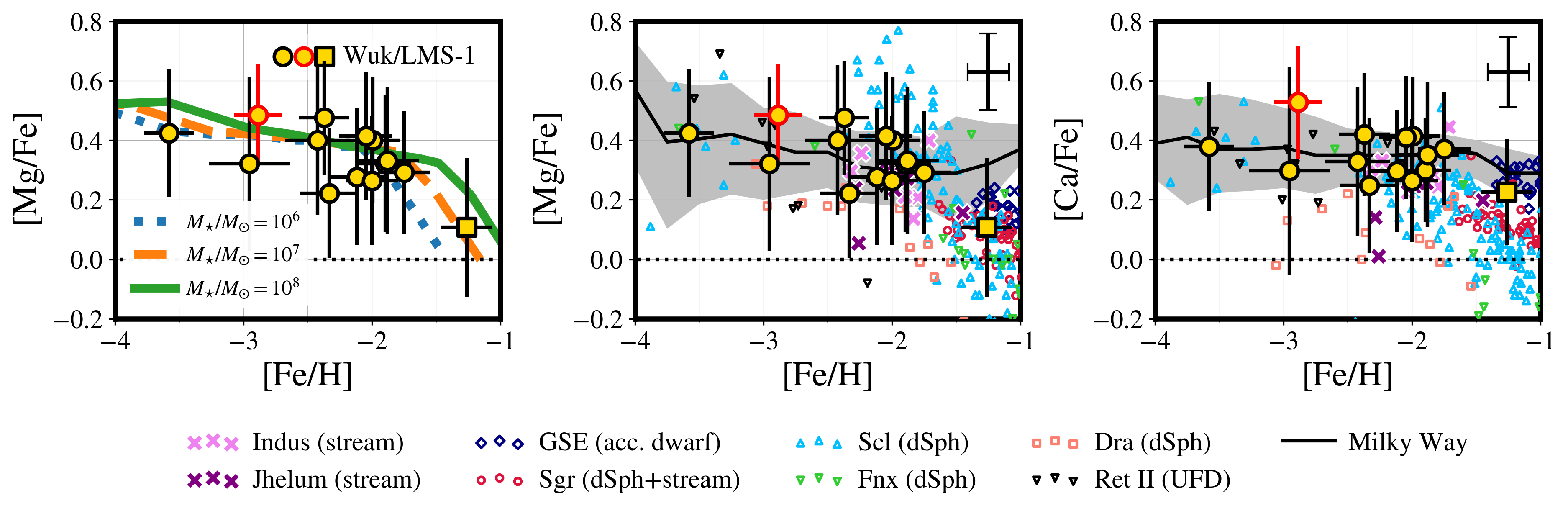}
\caption{Abundance trends of $\alpha$-elements Mg (left and middle) and Ca (right) in $\rm[\alpha/Fe]$ versus [Fe/H] format. Yellow symbols are our Wukong/LMS-1 stars. The marker with red edge represents the CEMP star in our sample (Wuk\_4). The yellow square is the most metal-rich star followed-up from APOGEE (Wuk\_14). Pink and purple crosses with white edges are stars from Indus and Jhelum stellar streams, respectively. In the left panel, blue (dotted), orange (dashed), and green (solid) lines are chemical evolution trajectories for galaxies with stellar masses $M_\star/M_\odot$ of $10^6$, $10^7$, and $10^8$, respectively \citep[][see text]{Wanajo2021}. In the middle and right panels, the black line shows the median abundances from the SAGA data base compilation of metal-poor stars in bins of 0.25\,dex in [Fe/H]. The grey band represents $16^{\rm th}$ and $84^{\rm th}$ percentiles within the same bins. Stars from Sagittarius (Sgr), Sculptor (Scl), Fornax (Fnx), and Draco (Dra) dSph galaxies, Reticulum II (Ret II) UFD, as well as GSE disrupted dwarf, are exhibited for comparison. See text for details on our abundance compilation for these dwarf galaxies. The solar abundance level is shown as dotted black lines in all panels. The black error bars in the top right corner of the middle and right panels show statistical uncertainties alone in [Mg/Fe] and [Ca/Fe], respectively, as well as [Fe/H].
\label{fig:alpha_abunds}}
\end{figure*}

In Figure \ref{fig:cemp_abunds}, we exhibit the same SAGA data base low-metallicity compilation from Figure \ref{fig:all_abunds}, but, now, dividing into carbon-normal (grey points), regular CEMP (orange diamonds), and CEMP-$s$ (blue triangles) stars. From the top panels, it becomes clear that CEMP-$s$ stars are also preferentially enhanced in Sr, Y, and Zr. Nevertheless, there are a few examples of ordinary CEMP stars with high values of [Sr/Fe], [Y/Fe], and/or [Zr/Fe] similar to Wuk\_4. 
In any case, it is also rather suspicious that Wuk\_4 is the only star with significant (${\approx}10$\kms, ${\gtrsim}5\sigma$) RV difference between H3's and our measurement (Section \ref{sec:rv}), possibly indicating a binary system. Indeed, binarity is believed to be the conventional pathway for the formation of CEMP-$s$ stars as they would experience mass transfer from an asymptotic giant branch companion (e.g., \citealt{Lucatello2005} and \citealt{Hansen2016cempNO} for RV monitoring studies and \citealt{Travaglio2004sprocess} for nucleosynthesis). Nevertheless, CEMP-no stars, those without any signatures of $s$-process enhancement ($\rm[Ba/Fe] < 0.0$) are also, sometimes, found in long period binaries \citep{Arenstsen2019cemp, Bonifacio2020cemp}. Deciphering whether or not mass transfer in a binary can create the excess of Sr, Y, and Zr in Wuk\_4 without high Ba will demand a systematic investigation of other CEMP stars with similar abundance patterns in the future. 


\subsection{\texorpdfstring{$\alpha$}x elements reveal a relatively massive dwarf galaxy
} \label{sec:alpha}

\begin{figure}
\centering
\includegraphics[width=1.0\columnwidth]{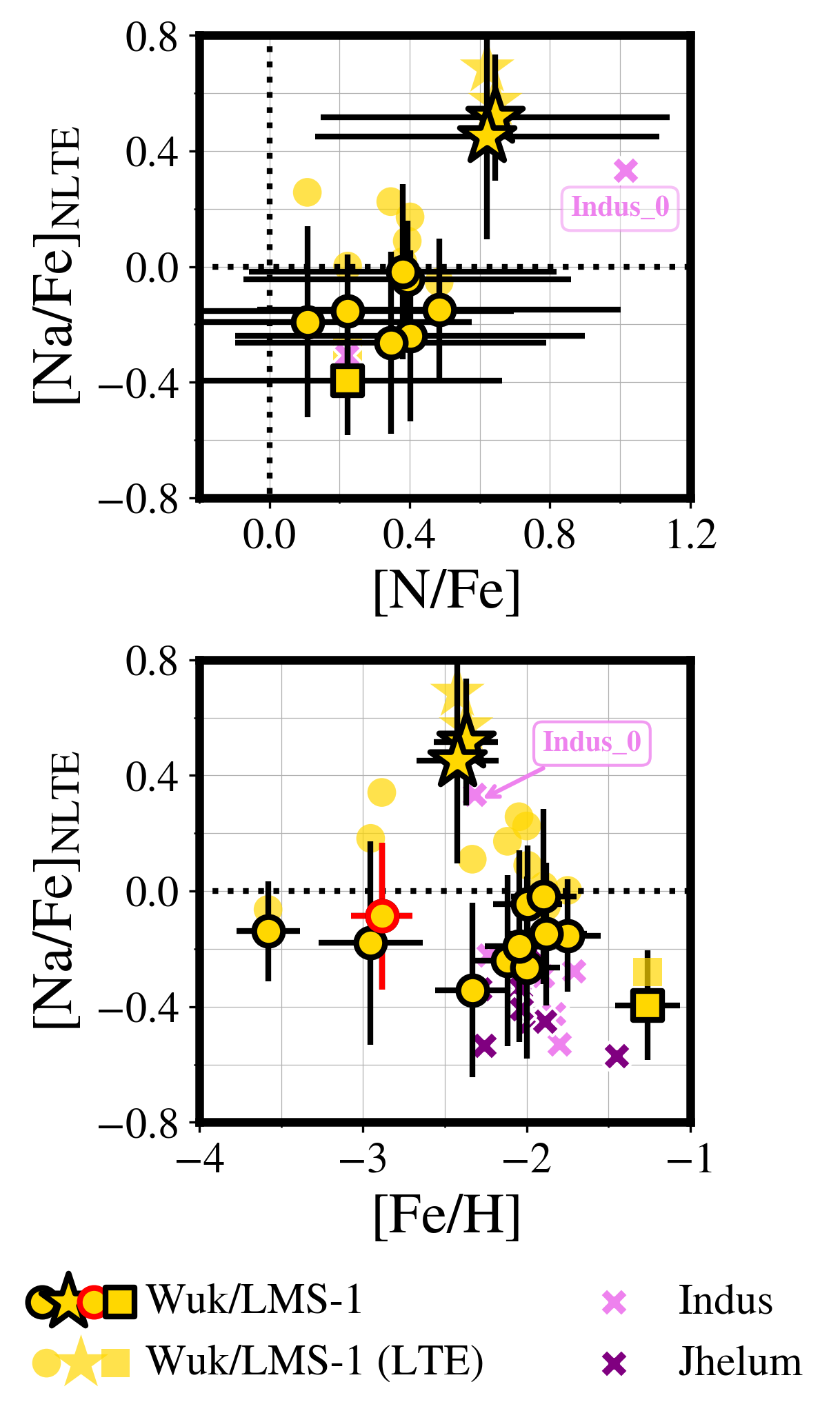}
\caption{Top: $\rm[Na/Fe]_{\rm NLTE}$ versus [N/Fe]. Bottom: $\rm[Na/Fe]_{\rm NLTE}$ versus [Fe/H]. In both panels, $\rm[Na/Fe]_{\rm NLTE}$ represents our sodium-to-iron ratios corrected with the NLTE departures from \citet[][see text]{Lind2011nlte_forNa}. Yellow symbols are Wukong/LMS-1 stars in our sample. The marker with red edge represents the CEMP star in our sample (Wuk\_4). The yellow square is the most metal-rich star followed-up from APOGEE (Wuk\_14). Star symbols represent the N-/Na-rich stars (Wuk\_5 and Wuk\_11). Transparent symbols without edges in the background correspond the same Wukong/LMS-1 stars, but without the NLTE corrections to Na. Pink and purple crosses with white edges are stars from Indus and Jhelum stellar streams, respectively. Indus\_0, which is also N-/Na-rich, is tagged in both panels. The solar abundance level is shown as dotted black lines in both panels.
\label{fig:NaFe_abunds}}
\end{figure}

\indent\indent Now, we look at the [$\alpha$/Fe] ratios in Wukong/LMS-1 stars, specifically the abundances of Mg and Ca. 
We only derived upper limits for O and Si is less reliable than Mg and Ca in the low-metallicity regime (see discussion in \citealt{Ji2020streams}). The behaviour of $\alpha$ abundances is a great tracer of the overall star-formation history of a galaxy \citep{Tinsley1979, Matteucci2012book}. More massive systems are expected to enrich themselves (i.e., reach higher metallicities) via core-collapse supernovae before the occurrence of Type Ia supernovae, which is delayed by ${\gtrsim}100$\,Myr \citep[e.g.,][]{Maoz2012dtdSNIa, delosReyes2022}. Because core-collapse supernovae 
produces mostly $\alpha$ elements while the Type Ia create almost exclusively iron-peak elements \citep[for a review, see][]{Nomoto2013}, a characteristic downturn in [$\alpha$/Fe] is expected at a certain metallicity (referred to as the ``knee''; \citealt{Matteucci1986}). The location of this knee in [Fe/H] depends on the star-formation and outflow efficiency, hence the mass, of a given galaxy (e.g., \citealt{Matteucci1990} and \citealt{Tolstoy2009}).
This is illustrated in Figure \ref{fig:alpha_abunds} (left panel) by the simulated chemical-evolution trajectories of \citet[][their figure 4, case 1]{Wanajo2021} for dwarf galaxies with different stellar masses ($10^6 \leq M_\star/M_\odot \leq 10^8$) at redshift $z=0$, similar to canonical dSph Milky Way satellites and encompassing the expected value for Wukong/LMS-1's progenitor. We note that we subtracted 0.25\,dex from their model's [Mg/H] values so the ``plateau'' is positioned at about $+0.4$\,dex. These recalibrated versions of the models will also be used in Section \ref{sec:rprocess}.

In Wukong/LMS-1, our derived abundances of both Mg and Ca reveal that the [$\alpha$/Fe] ratio remains high ($\sim$0.3--0.4\,dex) up to $\rm[Fe/H] \gtrsim -2$ (Figure \ref{fig:alpha_abunds}). This is, indeed, similar to somewhat massive surviving Milky Way satellites \citep[][]{Kirby2011alphas, Reichert2020alphaKnees}
. Before our analysis, the evidence for Wukong/LMS-1's progenitor being a relatively massive dwarf ($M_\star \sim 10^7 M_\odot$) came from tentative dynamical $N$-body modeling \citep{Malhan2021lms1}, the scaling relation between the total mass of its globular clusters and a galaxy's mass \citep[also][]{Malhan2021lms1}, or simple star counts \citep{Naidu2022mzr}. Hence, we provide the first evidence in favor of this hypothesis from chemistry. An identical $\alpha$-element abundance pattern is seen in Indus and Jhelum \citep[pink and purple crosses, respectively, in Figures \ref{fig:cemp_abunds} and beyond;][]{Ji2020streams}. As previously mentioned, although this chemical similarity between these streams does not fully confirm their association, it still corroborates such scenario.

Although the models in Figure \ref{fig:alpha_abunds} exemplify how a galaxy's mass correlates with its chemical evolution, we do not claim to be actually measuring the stellar mass of Wukong/LMS-1 from abundances. Systematics make the uncertainties of our [Mg/Fe] and [Ca/Fe] estimates to be typically 0.2\,dex. For reference, the median statistical uncertainties are $\approx$0.1\,dex for these [$\alpha$/Fe] ratios (top right corner in middle and right panels of Figure \ref{fig:alpha_abunds}). Not only this precision does not allow us to differentiate between those models, but the models themselves carry potentially even worse systematics such as supernovae and/or NSM yields. Nevertheless, we reinforce that the constant, within errors, [$\alpha$/Fe] up to $\rm[Fe/H] \sim -2$ is in conformity with the behaviour seen in the \textit{data} of nearby massive ($M_\star/M_\odot \geq 10^6$) dwarfs.

It is also relevant that the most metal-rich star ($\rm[Fe/H] = -1.26\pm0.20$, Wuk\_14; yellow square in all figures) in the sample has lower [Mg/Fe] and [Ca/Fe] than the bulk of our observed Wukong/LMS-1 members ($\rm[Fe/H] \leq -1.75$) by 0.1--0.2\,dex (middle and right panels of Figure \ref{fig:alpha_abunds}). This is, perhaps, not a particularly surprising result given that we observed this additional star specifically with the goal of testing the standard chemical enrichment scenario as discussed above. Nevertheless, we notice that Jhelum also contains a low-$\alpha$ star at $\rm[Fe/H] = -1.45$, which adds to the emerging picture where Wukong/LMS-1 progenitor experienced quite a simple chemical evolution pathway. It is also informative that Wukong/LMS-1 shows no evidence for additional bursts of star formation, which would cause the [$\alpha$/Fe] to actually increase at higher [Fe/H], a phenomenon that happens to some dwarfs, such as Sagittarius, Fornax, and the Magellanic Clouds (see \citealt{Nidever2020} and \citealt{Hasselquist2021dwarf_gals}), and is expected to be caused by their interaction with their massive host (in this case, the Milky Way). 
We did not find any trustworthy Wukong/LMS-1 candidates at metallicities higher than Wuk\_14 neither in H3 nor in APOGEE.


\subsection{A pair of stars with abundances reminiscent of multiple stellar populations in globular clusters} \label{sec:gc_stars}

\indent\indent Almost all globular clusters are known to contain more than just a single, sometimes several, well-defined sequences in optical/ultraviolet color-magnitude diagrams \citep{piotto2015, milone2017}. Color differences between these distinct tracks are caused by star-to-star variations in their abundances of light elements, including correlated enhancements in N and Na \citep[e.g.,][]{Gratton2004review}. Such phenomenon is referred to as ``multiple stellar populations'' \citep{BastianLardo2018mps, Milone2022mps}. Star clusters with multiple stellar populations have been detected not only in the Milky Way, but also other galaxies in the Local Group, including the Magellanic Clouds \citep{Mucciarelli2009lmc, Dalessandro2016smc} and even Fornax dSph \citep{Larsen2014fornax}. Most recently, evidence has been provided for multiple stellar populations in globular-cluster stellar streams \citep{Balbinot2022gd1, Martin2022c19, Usman2024300s}. Interestingly, a clear detection of a N-/Na-rich star was actually first made in Indus \citep{Ji2020streams}, which holds even after NLTE corrections to Na \citep{Hansen2021indus}. 

Wukong/LMS-1 contains an additional pair of N-/Na-rich stars (top panel of Figure \ref{fig:NaFe_abunds}), reinforcing the idea that its progenitor dwarf galaxy had not only NGC 5024 and NGC 5053 as globular clusters, but also at least one other that has been fully disrupted \citep{Hansen2021indus}, which is not unexpected for its mass \citep{Eadie2022gcs}. In order to guarantee that this result is robust against NLTE effects, we performed corrections following \citet{Lind2011nlte_forNa}. In practice, we corrected Na abundances line by line and, then, recomputed the weighted averages using our formalism (Section \ref{sec:abund}). In this process, uncertainties and weights remain unchanged. For consistency, we applied the same approach to both Indus and Jhelum stars. As can be appreciated from Figure \ref{fig:NaFe_abunds}, NLTE corrections to Na exacerbate the differences between the bulk of our Wukong/LMS-1 members and its enriched stars; original LTE abundances are plotted in the background as transparent symbols. We note that both NGC 5024 and NGC 5053 are also very metal-poor ($\rm[Fe/H] \lesssim -2$; \citealt{Kruijssen2019milkyway} and references therein) and contain multiple stellar populations \citep[see their ultraviolet color-magnitude diagrams in][]{piotto2015}

From the bottom panel of Figure \ref{fig:NaFe_abunds}, it is intriguing that both of our N-/Na-rich Wukong/LMS-1 stars (Wuk\_5 at $\rm[Fe/H] = -2.42\pm0.25$ and Wuk\_11 at $\rm[Fe/H] = -2.37\pm0.20$) have metallicities compatible ($1\sigma$) to their analog in Indus (Indus\_0; $\rm[Fe/H] = -2.32\pm0.22$). We speculate that this could indicate that all these stars originated from a single disrupted globular cluster, though there is still not enough evidence to make a clear association.
Nevertheless, if more stars at this metallicity of ${\approx}{-}2.4$\,dex are found to have excess of N and Na in Indus, Jhelum, and the large Wukong/LMS-1, the scenario where they all belonged to the same, now completely disrupted, globular cluster would be reinforced. Furthermore, this underscores the exciting possibility of identifying more enriched globular-cluster stars embedded within other dwarf-galaxy streams, which would be further evidence in favor of the multiple stellar populations phenomenon being ubiquitous across different environments.

Now that we have strong indication that Wukong/LMS-1 had ${\geq}3$ globular clusters, we can utilize this information as independent constraint on the total mass of this dwarf (e.g., \citealt{Forbes2020} and \citealt{Callingham2022gcs}). We employ the relationship between the total number of globular clusters in a galaxy and its halo virial mass from \citet[][their equation 1]{BurkertForbes2020}. This exercise gives a total mass of ${\approx}10^{10} M_\odot$ for Wukong/LMS-1. This value, indeed, translates into a stellar mass of ${\sim}10^7$ assuming the stellar-to-halo mass relation from \citet{Rodriguez-Puebla2017} at redshift $z=1$ (${\sim}8$\,Gyr ago in \citealt{PlanckCollab2020} cosmology), consistent with expectations for Wukong/LMS-1 \citep{Malhan2021lms1, Naidu2022mzr}. From this line of reasoning, Wukong/LMS-1 might have contributed with ${\sim}1$\% of the present-day total mass of the Milky Way \citep[e.g.,][]{BlandHawthorn2016}.

\subsection{The rise of the \texorpdfstring{$r$}x-process in Wukong/LMS-1} \label{sec:rprocess}

\indent\indent Finally, we look at the abundance patterns of rapid ($r$-) neutron capture process elements in Wukong/LMS-1. In 2017, the electromagnetic counterpart of the neutron-star merger (NSM) event GW170817 \citep{Abbott2017a, Abbott2017_B, Abbott2017c} provided confirmation that this site is capable of producing copious amounts of heavy elements via the $r$-process \citep{drout2017, Kilpatrick2017, shappee2017}. Notwithstanding, although NSMs could be the \textit{only} site for the $r$-process
, evidence has been presented that additional sources are involved \citep[see the review by][]{Cowan2021rprocess}
, in particular a prompt source might be needed to explain the full abundance patterns of both the Milky Way \citep{Cote2019, Haynes2019, Kobayashi2020, Tsujimoto2021rprocess} and some of its dSph satellites \citep{Skuladottir2019, Reichert2020alphaKnees, Skuladottir2020rprocess}. The timescales for $r$-process enrichment embedded into the chemical abundances of stars hold clues to its dominant astrophysical site(s).

\begin{figure}
\centering
\includegraphics[width=1.0\columnwidth]{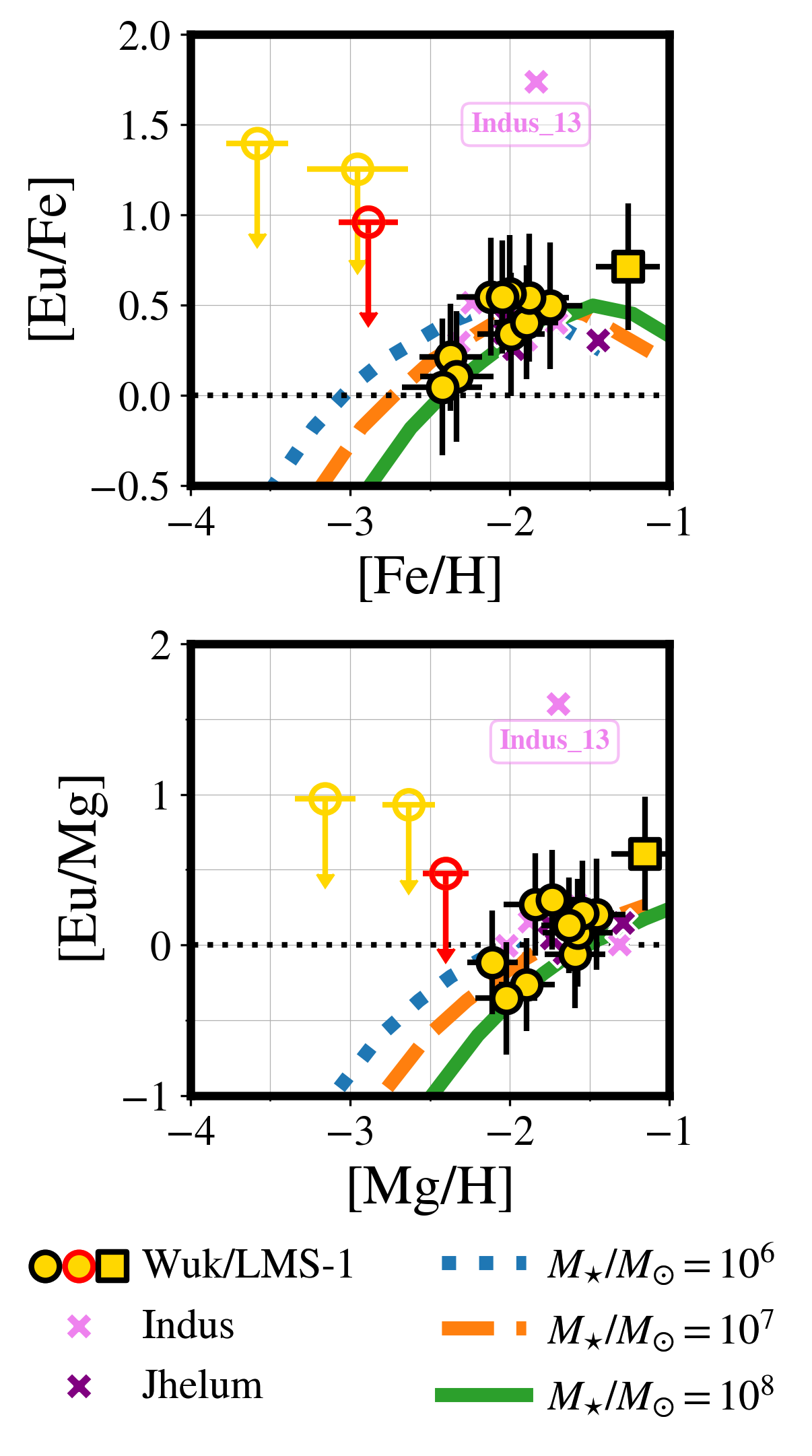}
\caption{Top: [Eu/Fe] versus [Fe/H]. Bottom: [Eu/Mg] versus [Mg/H].  Yellow symbols are Wukong/LMS-1 stars in our sample. Empty symbols are upper limits. The marker with red edge represents the CEMP star in our sample (Wuk\_4). The yellow square is the most metal-rich star followed-up from APOGEE (Wuk\_14). Pink and purple crosses with white edges are stars from Indus and Jhelum stellar streams, respectively. Indus\_13, which is an extremely $r$-process enhanced star is tagged in the top panel. In both panels, blue (dotted), orange (dashed), and green (solid) lines are chemical evolution trajectories for galaxies with stellar masses $M_\star/M_\odot$ of $10^6$, $10^7$, and $10^8$, respectively \citep[][see text]{Wanajo2021}. The solar abundance level is shown as dotted black lines in both panels.
\label{fig:r-process}}
\end{figure}

As can be seen in the top panel of Figure \ref{fig:r-process}, the [Eu/Fe] ratio, where Eu is mostly produced by the $r$-process \citep[see][]{Sneden2008}, increases from ${<}{+}0.1$\,dex at $\rm[Fe/H] \sim -2.4$ all the way up to ${>}{+}0.7$\,dex at $\rm[Fe/H] \sim -1.3$. The [Ba/Eu] ratio of Wukong/LMS-1 stars is effectively constant at ${\sim}{-}0.4$\,dex, confirming that the production of these elements is dominated by the $r$-process. For reference, a ``pure'' $r$-process signature is close to $\rm[Ba/Eu] = -0.8$ \citep[e.g.,][]{Bisterzo2014}. These abundances overlap with both Indus and Jhelum, although the data for these other stellar streams do not cover as large of a metallicity interval. We also recall the existence of the extremely $r$-process rich star Indus\_13 \citep{Hansen2021indus}, which is labeled in Figure \ref{fig:r-process}. The previously mentioned \citet{Wanajo2021} chemical-evolution sequences (Section \ref{sec:alpha}) are plotted in the background of Figure \ref{fig:r-process}. These models contain $r$-process production exclusively from NSMs. The current data for Wukong/LMS-1 seems compatible with this scenario, without the need for a prompt source (such as core-collapse supernovae) where [Eu/Fe] (or [Eu/Mg] as in the bottom panel of Figure \ref{fig:r-process}) would be constant over the low-metallicity regime. Note that we do not claim that NSMs are the only source of $r$-process in Wukong, but rather the dominant one within the metallicity range probed by our current data. 

An increasing trend of [Eu/Fe] with [Fe/H] has also been claimed to exist in UMi dSph \citep{Cohen2010}. However, this behaviour is accompanied by a huge scatter, which suggests that UMi actually experiences stochastic $r$-process enrichment, which, indeed, would be in line with its lower stellar mass ($M_\star \sim 10^5M_\odot$) in comparison to Sculptor, Fornax, and Sagittarius dSph galaxies ($M_\star > 10^6M_\odot$). Similar observations have also been made in another dwarf-galaxy stellar stream, the so-called ``Typhon'' \citep[][and \citealt{AlexJi2023typhon} for abundances]{Tenachi2022typhon, Dodd2023haloSubsGaiaDR3}. Nevertheless, the available data for that stream covers a shorter metallicity interval and the differences in [Eu/Fe] are smaller than in Wukong/LMS-1. Also, in Typhon, this result depends on a single star at $\rm[Fe/H] < -2$. Therefore, Wukong/LMS-1 constitutes the first example of a dwarf galaxy, though in the form of a stellar stream, with $r$-process enrichment that is clearly dominated by delayed sources, presumably NSMs. Of course, additional measurements of Eu at lower metallicities ($\rm[Fe/H]<-2.5$) will be necessary to test for the presence of a prompt $r$-process source in Wukong/LMS-1. Unfortunately, all stars analysed in this work at such low-metallicity regime are relatively hot ($\Teff \gtrsim 5100$\,K; Table \ref{tab:obs_mike}, Wuk\_4, 6, and 7), which impedes us from obtaining detections of Eu with the lines considered (4129\AA, 4205\AA, 4435\AA, 4522\AA, and 6645\AA).

\section{Conclusions} \label{sec:conclusions}

\indent\indent We presented results from the first spectroscopic follow-up campaign for stars in the Wukong/LMS-1 dwarf-galaxy stellar stream with the 
Magellan Clay/MIKE combination. These targets were originally identified as members from the H3 \citep{naidu2020, Johnson2022wukong} or APOGEE (Section \ref{sec:obs}) surveys. From these high-resolution MIKE spectra, we obtained RVs, stellar parameters, and detailed chemical abundances for 14 stars in Wukong/LMS-1 covering an extensive metallicity range ($-3.5 < \rm[Fe/H] \lesssim -1.3$). We also recalculated average abundances for a pair of stellar streams, Indus and Jhelum \citep{Ji2020streams}, which have been suggested to be associated with the larger Wukong/LMS-1 \citep{Bonaca2021streams, Malhan2021lms1, Malhan2022atlas}, in order to guarantee a consistent scale across samples. Our main results are summarized below.
\begin{itemize}
    \item Wukong/LMS-1 is chemically indistinguishable from Indus and Jhelum. Although this is not enough to confirm that they originated in the same parent dwarf galaxy, this is certainly evidence in favor of this hypothesis. 
    \item We confirmed a CEMP star in Wukong/LMS-1 (Wuk\_4; $\rm[Fe/H] = -2.89\pm0.19$ and $\rm[C/Fe] = +0.74\pm0.25$) with evidence for RV variation (${>}5\sigma$ level) as well as peculiar enhancements ($+0.5 \lesssim \rm[X/Fe] \leq +1.0$) in Sr, Y, and Zr, which is reminiscent of the CEMP-$s$ class, but without the expected high [Ba/Fe] ratio.
    \item The $\rm[\alpha/Fe]$ ratios (Mg, Ca) in Wukong/LMS-1 remain high ($\sim$0.3--0.4\,dex) up to $\rm[Fe/H] \gtrsim -2$, which is similar to relatively massive surviving dSph satellites of the Milky Way. This is in conformity with other works in the literature that estimated the mass of Wukong/LMS-1's progenitor with other methods \citep{Malhan2021lms1, Naidu2022mzr}. Moreover, the most metal-rich star in our sample (Wuk\_14; $\rm[Fe/H] = -1.26\pm0.20$), as well as a Jhelum star at $-1.45$\,dex, has lower [$\alpha$/Fe] by 0.1--0.2\,dex in comparison to the bulk of Wukong/LMS-1 stars, suggesting that this dwarf galaxy likely experienced fairly standard chemical evolution.
    \item Wukong/LMS-1 contains a pair of stars (Wuk\_5 and Wuk\_11) that are both N- and Na-rich in comparison to the bulk of the sample, which is a telltale sign that that these were born in a globular cluster with multiple stellar populations. This favors the hypothesis that Wukong/LMS-1 likely contained at least one globular cluster that has been completely disrupted. Interestingly, both of these Wukong/LMS-1 members plus a previously known N-/Na-rich Indus star all have compatible ($1\sigma$) metallicities ($\rm[Fe/H] \equiv 2.4$).
    \item Because Wukong/LMS-1 is also associated with a couple of intact globular clusters, NGC 5024 (M53) and NGC 5053, plus the disrupted one, we used this information to estimate the halo virial mass of the progenitor system. For ${\geq}3$ globular clusters, a galaxy is expected to have a total mass of ${\approx}10^{10}$, which corresponds to ${\sim}1$\% of the present-day Milky Way. 
    \item The [Eu/Fe] ratio in Wukong/LMS-1 stars increases as a function of metallicity within $-2.5 < \rm[Fe/H] \lesssim -1.3$, which can be reproduced by chemical-evolution models for similarly massive dwarfs with NSMs as the only source for the $r$-process \citep{Wanajo2021}, i.e., without the need for a prompt source. Wukong/LMS-1 is, in this context, the first example of dwarf galaxy with $r$-process enrichment clearly dominated by delayed sources. 
\end{itemize}

This paper provides a powerful demonstration of how detailed abundances can be used to unveil the evolution of disrupted dwarf galaxies, which, given their low masses and high accretion redshift, can not be spatially resolved, or detected (depending on the exact redshift), even by \textit{JWST}. The combination between \textit{Gaia} and complementary spectroscopic surveys can be employed to confidently identify members of streams/substructures if one is well-informed regarding potential interlopers, in particular from GSE and Sagittarius stream. We envision that the next step for Galactic archaeology will be to obtain samples covering larger metallicity ranges for these streams/substructures. Towards the low-metallicity regime, the number of available targets drastically diminishes. In the metal-rich end, Milky Way's halo foreground contamination is difficult to deal with. Therefore, dedicated searches for these targets will be necessary for us to continue advancing our knowledge about the fundamental building blocks of our Galaxy.

\section*{Acknowledgements}

\indent\indent This paper includes data gathered with the 6.5~meter Magellan Telescopes located at Las Campanas Observatory, Chile. This work benefited from discussions conducted during the ``Non-LTE workshop'' supported by the National Science Foundation under Grant No. OISE-1927130 (IReNA), the Kavli Institute for Cosmological Physics, and the University of Chicago Data Science Institute. G.L. is particularly thankful for discussions with Christian Hayes, Keith Hawkins, Kim Venn, and Vini Placco. G.L. is also indebted to \'Asa Sk\'ulad\'ottir for her inspiring talk, as well as afterward conversations, at the 2023 CeNAM Frontiers in Nuclear Astrophysics meeting. G.L. also thanks Zhen Yuan for providing the original LMS-1 members and Ting Li for pointing out the connection between Wukong/LMS-1 and Indus/Jhelum. G.L. is also sincerely grateful to Yutaka Hirai and Shinya Wanajo for providing the chemical-evolution models. 

We thank the anonymous referee whose comments and suggestions contributed to the quality of this paper. G.L. acknowledges FAPESP (procs. 2021/10429-0 and 2022/07301-5). A.P.J. acknowledges support by the National Science Foundation under grants AST-2206264 and AST-2307599. A.C. is supported by a Brinson Prize Fellowship at UChicago/KICP. S.R. thanks support from FAPESP (procs. 2015/50374-0 and 2020/15245-2), CAPES, and CNPq. Y.S.T. acknowledges financial support from the Australian Research Council through DECRA Fellowship DE220101520. L.B. acknowledges CAPES/PROEX (proc. 88887.821814/2023-00).

This work has made use of data from the European Space Agency (ESA) mission
{\it Gaia} (\url{https://www.cosmos.esa.int/gaia}), processed by the {\it Gaia}
Data Processing and Analysis Consortium (DPAC,
\url{https://www.cosmos.esa.int/web/gaia/dpac/consortium}). Funding for the DPAC
has been provided by national institutions, in particular the institutions
participating in the {\it Gaia} Multilateral Agreement.
This research has made use of NASA's Astrophysics Data System Bibliographic Services.

Funding for the Sloan Digital Sky Survey IV has been provided by the Alfred P. Sloan Foundation, the U.S. Department of Energy Office of Science, and the Participating Institutions. SDSS-IV acknowledges support and resources from the Center for High Performance Computing  at the University of Utah. The SDSS website is \url{www.sdss.org}. SDSS-IV is managed by the Astrophysical Research Consortium for the Participating Institutions of the SDSS Collaboration.

\section*{Data Availability}

\indent\indent The individual line measurements as well as full abundance tables are provided as supplementary material. All the data, including reduced spectra, plus full line measurements and abundance tables are available at \url{zenodo.org/records/10932808}. The code to generate complete abundance tables from \code{smhr}/\code{LESSPayne} outputs, as well as the redetermined abundances for Indus and Jhelum, can be found at \url{github.com/guilimberg/abund-tables}.



\bibliographystyle{mnras}
\bibliography{main}

\begin{thebibliography}{}
\makeatletter
\relax
\def\mn@urlcharsother{\let\do\@makeother \do\$\do\&\do\#\do\^\do\_\do\%\do\~}
\def\mn@doi{\begingroup\mn@urlcharsother \@ifnextchar [ {\mn@doi@}
  {\mn@doi@[]}}
\def\mn@doi@[#1]#2{\def\@tempa{#1}\ifx\@tempa\@empty \href
  {http://dx.doi.org/#2} {doi:#2}\else \href {http://dx.doi.org/#2} {#1}\fi
  \endgroup}
\def\mn@eprint#1#2{\mn@eprint@#1:#2::\@nil}
\def\mn@eprint@arXiv#1{\href {http://arxiv.org/abs/#1} {{\tt arXiv:#1}}}
\def\mn@eprint@dblp#1{\href {http://dblp.uni-trier.de/rec/bibtex/#1.xml}
  {dblp:#1}}
\def\mn@eprint@#1:#2:#3:#4\@nil{\def\@tempa {#1}\def\@tempb {#2}\def\@tempc
  {#3}\ifx \@tempc \@empty \let \@tempc \@tempb \let \@tempb \@tempa \fi \ifx
  \@tempb \@empty \def\@tempb {arXiv}\fi \@ifundefined
  {mn@eprint@\@tempb}{\@tempb:\@tempc}{\expandafter \expandafter \csname
  mn@eprint@\@tempb\endcsname \expandafter{\@tempc}}}

\bibitem[\protect\citeauthoryear{{Abbott} et~al.,}{{Abbott}
  et~al.}{2017a}]{Abbott2017a}
{Abbott} B.~P.,  et~al., 2017a, \mn@doi [\prl]
  {10.1103/PhysRevLett.119.161101}, \href
  {https://ui.adsabs.harvard.edu/abs/2017PhRvL.119p1101A} {119, 161101}

\bibitem[\protect\citeauthoryear{{Abbott} et~al.,}{{Abbott}
  et~al.}{2017b}]{Abbott2017_B}
{Abbott} B.~P.,  et~al., 2017b, \mn@doi [\apjl] {10.3847/2041-8213/aa91c9},
  \href {https://ui.adsabs.harvard.edu/abs/2017ApJ...848L..12A} {848, L12}

\bibitem[\protect\citeauthoryear{{Abbott} et~al.,}{{Abbott}
  et~al.}{2017c}]{Abbott2017c}
{Abbott} B.~P.,  et~al., 2017c, \mn@doi [\apjl] {10.3847/2041-8213/aa920c},
  \href {https://ui.adsabs.harvard.edu/abs/2017ApJ...848L..13A} {848, L13}

\bibitem[\protect\citeauthoryear{{Abdurro'uf} et~al.,}{{Abdurro'uf}
  et~al.}{2022}]{APOGEEdr17}
{Abdurro'uf} et~al., 2022, \mn@doi [\apjs] {10.3847/1538-4365/ac4414}, \href
  {https://ui.adsabs.harvard.edu/abs/2022ApJS..259...35A} {259, 35}

\bibitem[\protect\citeauthoryear{{Aguado} et~al.,}{{Aguado}
  et~al.}{2021a}]{Aguado2021}
{Aguado} D.~S.,  et~al., 2021a, \mn@doi [\mnras] {10.1093/mnras/staa3250},
  \href {https://ui.adsabs.harvard.edu/abs/2021MNRAS.500..889A} {500, 889}

\bibitem[\protect\citeauthoryear{{Aguado} et~al.,}{{Aguado}
  et~al.}{2021b}]{Aguado2021SausageSequoia}
{Aguado} D.~S.,  et~al., 2021b, \mn@doi [\apjl] {10.3847/2041-8213/abdbb8},
  \href {https://ui.adsabs.harvard.edu/abs/2021ApJ...908L...8A} {908, L8}

\bibitem[\protect\citeauthoryear{{Amarante}, {Debattista}, {Beraldo E Silva},
  {Laporte}  \& {Deg}}{{Amarante} et~al.}{2022}]{Amarante2022gsehalos}
{Amarante} J. A.~S.,  {Debattista} V.~P.,  {Beraldo E Silva} L.,  {Laporte} C.
  F.~P.,   {Deg} N.,  2022, \mn@doi [\apj] {10.3847/1538-4357/ac8b0d}, \href
  {https://ui.adsabs.harvard.edu/abs/2022ApJ...937...12A} {937, 12}

\bibitem[\protect\citeauthoryear{{Aoki}, {Beers}, {Christlieb}, {Norris},
  {Ryan}  \& {Tsangarides}}{{Aoki} et~al.}{2007}]{aoki2007}
{Aoki} W.,  {Beers} T.~C.,  {Christlieb} N.,  {Norris} J.~E.,  {Ryan} S.~G.,
  {Tsangarides} S.,  2007, \mn@doi [\apj] {10.1086/509817}, \href
  {https://ui.adsabs.harvard.edu/abs/2007ApJ...655..492A} {655, 492}

\bibitem[\protect\citeauthoryear{{Arentsen}, {Starkenburg}, {Shetrone}, {Venn},
  {Depagne}  \& {McConnachie}}{{Arentsen} et~al.}{2019}]{Arenstsen2019cemp}
{Arentsen} A.,  {Starkenburg} E.,  {Shetrone} M.~D.,  {Venn} K.~A.,  {Depagne}
  {\'E}.,   {McConnachie} A.~W.,  2019, \mn@doi [\aap]
  {10.1051/0004-6361/201834146}, \href
  {https://ui.adsabs.harvard.edu/abs/2019A&A...621A.108A} {621, A108}

\bibitem[\protect\citeauthoryear{{Arentsen}, {Placco}, {Lee}, {Aguado},
  {Martin}, {Starkenburg}  \& {Yoon}}{{Arentsen}
  et~al.}{2022}]{Arentsen2022cemp}
{Arentsen} A.,  {Placco} V.~M.,  {Lee} Y.~S.,  {Aguado} D.~S.,  {Martin} N.~F.,
   {Starkenburg} E.,   {Yoon} J.,  2022, arXiv e-prints, \href
  {https://ui.adsabs.harvard.edu/abs/2022arXiv220604081A} {p. arXiv:2206.04081}

\bibitem[\protect\citeauthoryear{{Asplund}}{{Asplund}}{2005}]{Asplund2005nlte}
{Asplund} M.,  2005, \mn@doi [\araa] {10.1146/annurev.astro.42.053102.134001},
  \href {https://ui.adsabs.harvard.edu/abs/2005ARA&A..43..481A} {43, 481}

\bibitem[\protect\citeauthoryear{{Asplund}, {Grevesse}, {Sauval}  \&
  {Scott}}{{Asplund} et~al.}{2009}]{asplund2009}
{Asplund} M.,  {Grevesse} N.,  {Sauval} A.~J.,   {Scott} P.,  2009, \mn@doi
  [\araa] {10.1146/annurev.astro.46.060407.145222}, \href
  {https://ui.adsabs.harvard.edu/abs/2009ARA&A..47..481A} {47, 481}

\bibitem[\protect\citeauthoryear{{Balbinot}, {Cabrera-Ziri}  \&
  {Lardo}}{{Balbinot} et~al.}{2022}]{Balbinot2022gd1}
{Balbinot} E.,  {Cabrera-Ziri} I.,   {Lardo} C.,  2022, \mn@doi [\mnras]
  {10.1093/mnras/stac1953}, \href
  {https://ui.adsabs.harvard.edu/abs/2022MNRAS.515.5802B} {515, 5802}

\bibitem[\protect\citeauthoryear{{Barklem} et~al.,}{{Barklem}
  et~al.}{2005}]{barklem2005}
{Barklem} P.~S.,  et~al., 2005, \mn@doi [\aap] {10.1051/0004-6361:20052967},
  \href {https://ui.adsabs.harvard.edu/abs/2005A&A...439..129B} {439, 129}

\bibitem[\protect\citeauthoryear{{Bastian} \& {Lardo}}{{Bastian} \&
  {Lardo}}{2018}]{BastianLardo2018mps}
{Bastian} N.,  {Lardo} C.,  2018, \mn@doi [\araa]
  {10.1146/annurev-astro-081817-051839}, \href
  {https://ui.adsabs.harvard.edu/abs/2018ARA&A..56...83B} {56, 83}

\bibitem[\protect\citeauthoryear{{Beers} \& {Christlieb}}{{Beers} \&
  {Christlieb}}{2005}]{beers2005}
{Beers} T.~C.,  {Christlieb} N.,  2005, \mn@doi [\araa]
  {10.1146/annurev.astro.42.053102.134057}, \href
  {https://ui.adsabs.harvard.edu/abs/2005ARA&A..43..531B} {43, 531}

\bibitem[\protect\citeauthoryear{{Beers}, {Preston}  \& {Shectman}}{{Beers}
  et~al.}{1992}]{beers1992}
{Beers} T.~C.,  {Preston} G.~W.,   {Shectman} S.~A.,  1992, AJ, 103, 1987

\bibitem[\protect\citeauthoryear{{Belokurov} et~al.,}{{Belokurov}
  et~al.}{2006}]{Belokurov2006Streams}
{Belokurov} V.,  et~al., 2006, \mn@doi [\apjl] {10.1086/504797}, \href
  {https://ui.adsabs.harvard.edu/abs/2006ApJ...642L.137B} {642, L137}

\bibitem[\protect\citeauthoryear{{Belokurov} et~al.,}{{Belokurov}
  et~al.}{2007}]{Belokurov2007Orphan}
{Belokurov} V.,  et~al., 2007, \mn@doi [\apj] {10.1086/511302}, \href
  {https://ui.adsabs.harvard.edu/abs/2007ApJ...658..337B} {658, 337}

\bibitem[\protect\citeauthoryear{{Belokurov}, {Erkal}, {Evans}, {Koposov}  \&
  {Deason}}{{Belokurov} et~al.}{2018}]{belokurov2018}
{Belokurov} V.,  {Erkal} D.,  {Evans} N.~W.,  {Koposov} S.~E.,   {Deason}
  A.~J.,  2018, \mn@doi [\mnras] {10.1093/mnras/sty982}, \href
  {https://ui.adsabs.harvard.edu/abs/2018MNRAS.478..611B} {478, 611}

\bibitem[\protect\citeauthoryear{{Bernstein}, {Shectman}, {Gunnels},
  {Mochnacki}  \& {Athey}}{{Bernstein} et~al.}{2003}]{Bernstein2003mike}
{Bernstein} R.,  {Shectman} S.~A.,  {Gunnels} S.~M.,  {Mochnacki} S.,   {Athey}
  A.~E.,  2003, in {Iye} M.,  {Moorwood} A. F.~M.,  eds,  Society of
  Photo-Optical Instrumentation Engineers (SPIE) Conference Series Vol. 4841,
  Instrument Design and Performance for Optical/Infrared Ground-based
  Telescopes. pp 1694--1704, \mn@doi{10.1117/12.461502}

\bibitem[\protect\citeauthoryear{{Bisterzo}, {Travaglio}, {Gallino}, {Wiescher}
   \& {K{\"a}ppeler}}{{Bisterzo} et~al.}{2014}]{Bisterzo2014}
{Bisterzo} S.,  {Travaglio} C.,  {Gallino} R.,  {Wiescher} M.,   {K{\"a}ppeler}
  F.,  2014, \mn@doi [\apj] {10.1088/0004-637X/787/1/10}, \href
  {https://ui.adsabs.harvard.edu/abs/2014ApJ...787...10B} {787, 10}

\bibitem[\protect\citeauthoryear{{Bland-Hawthorn} \&
  {Gerhard}}{{Bland-Hawthorn} \& {Gerhard}}{2016}]{BlandHawthorn2016}
{Bland-Hawthorn} J.,  {Gerhard} O.,  2016, \mn@doi [\araa]
  {10.1146/annurev-astro-081915-023441}, \href
  {https://ui.adsabs.harvard.edu/abs/2016ARA&A..54..529B} {54, 529}

\bibitem[\protect\citeauthoryear{{Bonaca}, {Conroy}, {Price-Whelan}  \&
  {Hogg}}{{Bonaca} et~al.}{2019}]{Bonaca2019indus+jhelum}
{Bonaca} A.,  {Conroy} C.,  {Price-Whelan} A.~M.,   {Hogg} D.~W.,  2019,
  \mn@doi [\apjl] {10.3847/2041-8213/ab36ba}, \href
  {https://ui.adsabs.harvard.edu/abs/2019ApJ...881L..37B} {881, L37}

\bibitem[\protect\citeauthoryear{{Bonaca} et~al.,}{{Bonaca}
  et~al.}{2021}]{Bonaca2021streams}
{Bonaca} A.,  et~al., 2021, \mn@doi [\apjl] {10.3847/2041-8213/abeaa9}, \href
  {https://ui.adsabs.harvard.edu/abs/2021ApJ...909L..26B} {909, L26}

\bibitem[\protect\citeauthoryear{{Bonifacio} et~al.,}{{Bonifacio}
  et~al.}{2020}]{Bonifacio2020cemp}
{Bonifacio} P.,  et~al., 2020, \mn@doi [\aap] {10.1051/0004-6361/201935833},
  \href {https://ui.adsabs.harvard.edu/abs/2020A&A...633A.129B} {633, A129}

\bibitem[\protect\citeauthoryear{{Boylan-Kolchin}, {Weisz}, {Johnson},
  {Bullock}, {Conroy}  \& {Fitts}}{{Boylan-Kolchin}
  et~al.}{2015}]{Boylan-Kolchin2015nearfield}
{Boylan-Kolchin} M.,  {Weisz} D.~R.,  {Johnson} B.~D.,  {Bullock} J.~S.,
  {Conroy} C.,   {Fitts} A.,  2015, \mn@doi [\mnras] {10.1093/mnras/stv1736},
  \href {https://ui.adsabs.harvard.edu/abs/2015MNRAS.453.1503B} {453, 1503}

\bibitem[\protect\citeauthoryear{{Boylan-Kolchin}, {Weisz}, {Bullock}  \&
  {Cooper}}{{Boylan-Kolchin} et~al.}{2016}]{Boylan-Kolchin2016nearfield}
{Boylan-Kolchin} M.,  {Weisz} D.~R.,  {Bullock} J.~S.,   {Cooper} M.~C.,  2016,
  \mn@doi [\mnras] {10.1093/mnrasl/slw121}, \href
  {https://ui.adsabs.harvard.edu/abs/2016MNRAS.462L..51B} {462, L51}

\bibitem[\protect\citeauthoryear{{Buder} et~al.,}{{Buder}
  et~al.}{2022}]{Buder2022halo}
{Buder} S.,  et~al., 2022, \mn@doi [\mnras] {10.1093/mnras/stab3504}, \href
  {https://ui.adsabs.harvard.edu/abs/2022MNRAS.510.2407B} {510, 2407}

\bibitem[\protect\citeauthoryear{{Bullock} \& {Johnston}}{{Bullock} \&
  {Johnston}}{2005}]{BullockJohnston2005}
{Bullock} J.~S.,  {Johnston} K.~V.,  2005, \mn@doi [\apj] {10.1086/497422},
  \href {https://ui.adsabs.harvard.edu/abs/2005ApJ...635..931B} {635, 931}

\bibitem[\protect\citeauthoryear{{Burkert} \& {Forbes}}{{Burkert} \&
  {Forbes}}{2020}]{BurkertForbes2020}
{Burkert} A.,  {Forbes} D.~A.,  2020, \mn@doi [\aj] {10.3847/1538-3881/ab5b0e},
  \href {https://ui.adsabs.harvard.edu/abs/2020AJ....159...56B} {159, 56}

\bibitem[\protect\citeauthoryear{{Callingham}, {Cautun}, {Deason}, {Frenk},
  {Grand}  \& {Marinacci}}{{Callingham} et~al.}{2022}]{Callingham2022gcs}
{Callingham} T.~M.,  {Cautun} M.,  {Deason} A.~J.,  {Frenk} C.~S.,  {Grand} R.
  J.~J.,   {Marinacci} F.,  2022, \mn@doi [\mnras] {10.1093/mnras/stac1145},
  \href {https://ui.adsabs.harvard.edu/abs/2022MNRAS.513.4107C} {513, 4107}

\bibitem[\protect\citeauthoryear{{Cargile}, {Conroy}, {Johnson}, {Ting},
  {Bonaca}, {Dotter}  \& {Speagle}}{{Cargile}
  et~al.}{2020}]{Cargile2020minesweeper}
{Cargile} P.~A.,  {Conroy} C.,  {Johnson} B.~D.,  {Ting} Y.-S.,  {Bonaca} A.,
  {Dotter} A.,   {Speagle} J.~S.,  2020, \mn@doi [\apj]
  {10.3847/1538-4357/aba43b}, \href
  {https://ui.adsabs.harvard.edu/abs/2020ApJ...900...28C} {900, 28}

\bibitem[\protect\citeauthoryear{{Casey}}{{Casey}}{2014}]{Casey2014smhr}
{Casey} A.~R.,  2014, PhD thesis, Australian National University, Canberra

\bibitem[\protect\citeauthoryear{{Casey}, {Keller}, {Da Costa}, {Frebel}  \&
  {Maunder}}{{Casey} et~al.}{2014}]{Casey2014orphan}
{Casey} A.~R.,  {Keller} S.~C.,  {Da Costa} G.,  {Frebel} A.,   {Maunder} E.,
  2014, \mn@doi [\apj] {10.1088/0004-637X/784/1/19}, \href
  {https://ui.adsabs.harvard.edu/abs/2014ApJ...784...19C} {784, 19}

\bibitem[\protect\citeauthoryear{{Castelli} \& {Kurucz}}{{Castelli} \&
  {Kurucz}}{2003}]{Castelli2003atmospheres}
{Castelli} F.,  {Kurucz} R.~L.,  2003, in {Piskunov} N.,  {Weiss} W.~W.,
  {Gray} D.~F.,  eds, ~ Vol. 210, Modelling of Stellar Atmospheres. p.~A20
  (\mn@eprint {arXiv} {astro-ph/0405087})

\bibitem[\protect\citeauthoryear{{Chiti} \& {Frebel}}{{Chiti} \&
  {Frebel}}{2019}]{Chiti2019sgr}
{Chiti} A.,  {Frebel} A.,  2019, \mn@doi [\apj] {10.3847/1538-4357/ab0f9f},
  \href {https://ui.adsabs.harvard.edu/abs/2019ApJ...875..112C} {875, 112}

\bibitem[\protect\citeauthoryear{{Chiti}, {Frebel}, {Ji}, {Jerjen}, {Kim}  \&
  {Norris}}{{Chiti} et~al.}{2018}]{chiti2018}
{Chiti} A.,  {Frebel} A.,  {Ji} A.~P.,  {Jerjen} H.,  {Kim} D.,   {Norris}
  J.~E.,  2018, \mn@doi [\apj] {10.3847/1538-4357/aab4fc}, \href
  {https://ui.adsabs.harvard.edu/abs/2018ApJ...857...74C} {857, 74}

\bibitem[\protect\citeauthoryear{{Chiti}, {Hansen}  \& {Frebel}}{{Chiti}
  et~al.}{2020}]{Chiti2020sgr}
{Chiti} A.,  {Hansen} K.~Y.,   {Frebel} A.,  2020, \mn@doi [\apj]
  {10.3847/1538-4357/abb1ae}, \href
  {https://ui.adsabs.harvard.edu/abs/2020ApJ...901..164C} {901, 164}

\bibitem[\protect\citeauthoryear{{Cohen} \& {Huang}}{{Cohen} \&
  {Huang}}{2009}]{Cohen2009}
{Cohen} J.~G.,  {Huang} W.,  2009, \mn@doi [\apj]
  {10.1088/0004-637X/701/2/1053}, \href
  {https://ui.adsabs.harvard.edu/abs/2009ApJ...701.1053C} {701, 1053}

\bibitem[\protect\citeauthoryear{{Cohen} \& {Huang}}{{Cohen} \&
  {Huang}}{2010}]{Cohen2010}
{Cohen} J.~G.,  {Huang} W.,  2010, \mn@doi [\apj]
  {10.1088/0004-637X/719/1/931}, \href
  {https://ui.adsabs.harvard.edu/abs/2010ApJ...719..931C} {719, 931}

\bibitem[\protect\citeauthoryear{{Cohen}, {Christlieb}, {Thompson},
  {McWilliam}, {Shectman}, {Reimers}, {Wisotzki}  \& {Kirby}}{{Cohen}
  et~al.}{2013}]{Cohen2013}
{Cohen} J.~G.,  {Christlieb} N.,  {Thompson} I.,  {McWilliam} A.,  {Shectman}
  S.,  {Reimers} D.,  {Wisotzki} L.,   {Kirby} E.,  2013, \mn@doi [\apj]
  {10.1088/0004-637X/778/1/56}, \href
  {https://ui.adsabs.harvard.edu/abs/2013ApJ...778...56C} {778, 56}

\bibitem[\protect\citeauthoryear{{Conroy} et~al.,}{{Conroy}
  et~al.}{2019}]{Conroy2019surveyH3}
{Conroy} C.,  et~al., 2019, \mn@doi [\apj] {10.3847/1538-4357/ab38b8}, \href
  {https://ui.adsabs.harvard.edu/abs/2019ApJ...883..107C} {883, 107}

\bibitem[\protect\citeauthoryear{{Cooper} et~al.,}{{Cooper}
  et~al.}{2010}]{Cooper2010}
{Cooper} A.~P.,  et~al., 2010, \mn@doi [\mnras]
  {10.1111/j.1365-2966.2010.16740.x}, \href
  {https://ui.adsabs.harvard.edu/abs/2010MNRAS.406..744C} {406, 744}

\bibitem[\protect\citeauthoryear{{Cooper}, {D'Souza}, {Kauffmann}, {Wang},
  {Boylan-Kolchin}, {Guo}, {Frenk}  \& {White}}{{Cooper}
  et~al.}{2013}]{Cooper2013}
{Cooper} A.~P.,  {D'Souza} R.,  {Kauffmann} G.,  {Wang} J.,  {Boylan-Kolchin}
  M.,  {Guo} Q.,  {Frenk} C.~S.,   {White} S. D.~M.,  2013, \mn@doi [\mnras]
  {10.1093/mnras/stt1245}, \href
  {https://ui.adsabs.harvard.edu/abs/2013MNRAS.434.3348C} {434, 3348}

\bibitem[\protect\citeauthoryear{{C{\^o}t{\'e}} et~al.,}{{C{\^o}t{\'e}}
  et~al.}{2019}]{Cote2019}
{C{\^o}t{\'e}} B.,  et~al., 2019, \mn@doi [\apj] {10.3847/1538-4357/ab10db},
  \href {https://ui.adsabs.harvard.edu/abs/2019ApJ...875..106C} {875, 106}

\bibitem[\protect\citeauthoryear{{Cowan}, {Sneden}, {Lawler}, {Aprahamian},
  {Wiescher}, {Langanke}, {Mart{\'\i}nez-Pinedo}  \& {Thielemann}}{{Cowan}
  et~al.}{2021}]{Cowan2021rprocess}
{Cowan} J.~J.,  {Sneden} C.,  {Lawler} J.~E.,  {Aprahamian} A.,  {Wiescher} M.,
   {Langanke} K.,  {Mart{\'\i}nez-Pinedo} G.,   {Thielemann} F.-K.,  2021,
  \mn@doi [Reviews of Modern Physics] {10.1103/RevModPhys.93.015002}, \href
  {https://ui.adsabs.harvard.edu/abs/2021RvMP...93a5002C} {93, 015002}

\bibitem[\protect\citeauthoryear{{Dalessandro}, {Lapenna}, {Mucciarelli},
  {Origlia}, {Ferraro}  \& {Lanzoni}}{{Dalessandro}
  et~al.}{2016}]{Dalessandro2016smc}
{Dalessandro} E.,  {Lapenna} E.,  {Mucciarelli} A.,  {Origlia} L.,  {Ferraro}
  F.~R.,   {Lanzoni} B.,  2016, \mn@doi [\apj] {10.3847/0004-637X/829/2/77},
  \href {https://ui.adsabs.harvard.edu/abs/2016ApJ...829...77D} {829, 77}

\bibitem[\protect\citeauthoryear{{Dodd}, {Callingham}, {Helmi}, {Matsuno},
  {Ruiz-Lara}, {Balbinot}  \& {L{\"o}vdal}}{{Dodd}
  et~al.}{2023}]{Dodd2023haloSubsGaiaDR3}
{Dodd} E.,  {Callingham} T.~M.,  {Helmi} A.,  {Matsuno} T.,  {Ruiz-Lara} T.,
  {Balbinot} E.,   {L{\"o}vdal} S.,  2023, \mn@doi [\aap]
  {10.1051/0004-6361/202244546}, \href
  {https://ui.adsabs.harvard.edu/abs/2023A&A...670L...2D} {670, L2}

\bibitem[\protect\citeauthoryear{{Drout} et~al.,}{{Drout}
  et~al.}{2017}]{drout2017}
{Drout} M.~R.,  et~al., 2017, \mn@doi [Science] {10.1126/science.aaq0049},
  \href {https://ui.adsabs.harvard.edu/abs/2017Sci...358.1570D} {358, 1570}

\bibitem[\protect\citeauthoryear{{Eadie}, {Harris}  \& {Springford}}{{Eadie}
  et~al.}{2022}]{Eadie2022gcs}
{Eadie} G.~M.,  {Harris} W.~E.,   {Springford} A.,  2022, \mn@doi [\apj]
  {10.3847/1538-4357/ac33b0}, \href
  {https://ui.adsabs.harvard.edu/abs/2022ApJ...926..162E} {926, 162}

\bibitem[\protect\citeauthoryear{{Ezzeddine} et~al.,}{{Ezzeddine}
  et~al.}{2020}]{ezzedine2020}
{Ezzeddine} R.,  et~al., 2020, \mn@doi [\apj] {10.3847/1538-4357/ab9d1a}, \href
  {https://ui.adsabs.harvard.edu/abs/2020ApJ...898..150E} {898, 150}

\bibitem[\protect\citeauthoryear{{Faber} \& {Gallagher}}{{Faber} \&
  {Gallagher}}{1979}]{FaberGallagher1979}
{Faber} S.~M.,  {Gallagher} J.~S.,  1979, \mn@doi [\araa]
  {10.1146/annurev.aa.17.090179.001031}, \href
  {https://ui.adsabs.harvard.edu/abs/1979ARA&A..17..135F} {17, 135}

\bibitem[\protect\citeauthoryear{{Forbes}}{{Forbes}}{2020}]{Forbes2020}
{Forbes} D.~A.,  2020, \mn@doi [\mnras] {10.1093/mnras/staa245}, \href
  {https://ui.adsabs.harvard.edu/abs/2020MNRAS.493..847F} {493, 847}

\bibitem[\protect\citeauthoryear{{Frebel}}{{Frebel}}{2018}]{frebel2018}
{Frebel} A.,  2018, \mn@doi [Annual Review of Nuclear and Particle Science]
  {10.1146/annurev-nucl-101917-021141}, \href
  {https://ui.adsabs.harvard.edu/abs/2018ARNPS..68..237F} {68, 237}

\bibitem[\protect\citeauthoryear{{Frebel} \& {Norris}}{{Frebel} \&
  {Norris}}{2015}]{frebel2015}
{Frebel} A.,  {Norris} J.~E.,  2015, \mn@doi [\araa]
  {10.1146/annurev-astro-082214-122423}, \href
  {https://ui.adsabs.harvard.edu/abs/2015ARA&A..53..631F} {53, 631}

\bibitem[\protect\citeauthoryear{{Frebel}, {Casey}, {Jacobson}  \&
  {Yu}}{{Frebel} et~al.}{2013}]{Frebel2013photscale}
{Frebel} A.,  {Casey} A.~R.,  {Jacobson} H.~R.,   {Yu} Q.,  2013, \mn@doi
  [\apj] {10.1088/0004-637X/769/1/57}, \href
  {https://ui.adsabs.harvard.edu/abs/2013ApJ...769...57F} {769, 57}

\bibitem[\protect\citeauthoryear{{Fulbright}}{{Fulbright}}{2000}]{Fulbright2000}
{Fulbright} J.~P.,  2000, \mn@doi [\aj] {10.1086/301548}, \href
  {https://ui.adsabs.harvard.edu/abs/2000AJ....120.1841F} {120, 1841}

\bibitem[\protect\citeauthoryear{{Gaia Collaboration} et~al.,}{{Gaia
  Collaboration} et~al.}{2016a}]{GaiaMission}
{Gaia Collaboration} et~al., 2016a, \mn@doi [\aap]
  {10.1051/0004-6361/201629272}, \href
  {https://ui.adsabs.harvard.edu/abs/2016A&A...595A...1G} {595, A1}

\bibitem[\protect\citeauthoryear{{Gaia Collaboration} et~al.,}{{Gaia
  Collaboration} et~al.}{2016b}]{gaiadr1}
{Gaia Collaboration} et~al., 2016b, \mn@doi [\aap]
  {10.1051/0004-6361/201629512}, \href
  {https://ui.adsabs.harvard.edu/abs/2016A&A...595A...2G} {595, A2}

\bibitem[\protect\citeauthoryear{{Gaia Collaboration} et~al.,}{{Gaia
  Collaboration} et~al.}{2018}]{gaiadr2}
{Gaia Collaboration} et~al., 2018, \mn@doi [\aap]
  {10.1051/0004-6361/201833051}, \href
  {https://ui.adsabs.harvard.edu/abs/2018A&A...616A...1G} {616, A1}

\bibitem[\protect\citeauthoryear{{Gaia Collaboration} et~al.,}{{Gaia
  Collaboration} et~al.}{2021}]{GaiaEDR3Summary}
{Gaia Collaboration} et~al., 2021, \mn@doi [\aap]
  {10.1051/0004-6361/202039657}, \href
  {https://ui.adsabs.harvard.edu/abs/2021A&A...649A...1G} {649, A1}

\bibitem[\protect\citeauthoryear{{Gaia Collaboration} et~al.,}{{Gaia
  Collaboration} et~al.}{2022}]{GaiaDR32022arXiv}
{Gaia Collaboration} et~al., 2022, arXiv e-prints, \href
  {https://ui.adsabs.harvard.edu/abs/2022arXiv220800211G} {p. arXiv:2208.00211}

\bibitem[\protect\citeauthoryear{{Geisler}, {Smith}, {Wallerstein}, {Gonzalez}
  \& {Charbonnel}}{{Geisler} et~al.}{2005}]{Geisler2005}
{Geisler} D.,  {Smith} V.~V.,  {Wallerstein} G.,  {Gonzalez} G.,   {Charbonnel}
  C.,  2005, \mn@doi [\aj] {10.1086/427540}, \href
  {https://ui.adsabs.harvard.edu/abs/2005AJ....129.1428G} {129, 1428}

\bibitem[\protect\citeauthoryear{{Gratton}, {Sneden}  \& {Carretta}}{{Gratton}
  et~al.}{2004}]{Gratton2004review}
{Gratton} R.,  {Sneden} C.,   {Carretta} E.,  2004, \mn@doi [\araa]
  {10.1146/annurev.astro.42.053102.133945}, \href
  {https://ui.adsabs.harvard.edu/abs/2004ARA&A..42..385G} {42, 385}

\bibitem[\protect\citeauthoryear{{Hansen}, {Andersen}, {Nordstr{\"o}m},
  {Beers}, {Placco}, {Yoon}  \& {Buchhave}}{{Hansen}
  et~al.}{2016}]{Hansen2016cempNO}
{Hansen} T.~T.,  {Andersen} J.,  {Nordstr{\"o}m} B.,  {Beers} T.~C.,  {Placco}
  V.~M.,  {Yoon} J.,   {Buchhave} L.~A.,  2016, \mn@doi [\aap]
  {10.1051/0004-6361/201527235}, \href
  {https://ui.adsabs.harvard.edu/abs/2016A&A...586A.160H} {586, A160}

\bibitem[\protect\citeauthoryear{{Hansen} et~al.,}{{Hansen}
  et~al.}{2021}]{Hansen2021indus}
{Hansen} T.~T.,  et~al., 2021, \mn@doi [\apj] {10.3847/1538-4357/abfc54}, \href
  {https://ui.adsabs.harvard.edu/abs/2021ApJ...915..103H} {915, 103}

\bibitem[\protect\citeauthoryear{{Hasselquist} et~al.,}{{Hasselquist}
  et~al.}{2019}]{Hasselquist2019sgr}
{Hasselquist} S.,  et~al., 2019, \mn@doi [\apj] {10.3847/1538-4357/aafdac},
  \href {https://ui.adsabs.harvard.edu/abs/2019ApJ...872...58H} {872, 58}

\bibitem[\protect\citeauthoryear{{Hasselquist} et~al.,}{{Hasselquist}
  et~al.}{2021}]{Hasselquist2021dwarf_gals}
{Hasselquist} S.,  et~al., 2021, \mn@doi [\apj] {10.3847/1538-4357/ac25f9},
  \href {https://ui.adsabs.harvard.edu/abs/2021ApJ...923..172H} {923, 172}

\bibitem[\protect\citeauthoryear{{Hawkins} et~al.,}{{Hawkins}
  et~al.}{2023}]{Hawkins2023orphan}
{Hawkins} K.,  et~al., 2023, \mn@doi [\apj] {10.3847/1538-4357/acb698}, \href
  {https://ui.adsabs.harvard.edu/abs/2023ApJ...948..123H} {948, 123}

\bibitem[\protect\citeauthoryear{{Hayes} et~al.,}{{Hayes}
  et~al.}{2020}]{Hayes2020}
{Hayes} C.~R.,  et~al., 2020, \mn@doi [\apj] {10.3847/1538-4357/ab62ad}, \href
  {https://ui.adsabs.harvard.edu/abs/2020ApJ...889...63H} {889, 63}

\bibitem[\protect\citeauthoryear{{Haynes} \& {Kobayashi}}{{Haynes} \&
  {Kobayashi}}{2019}]{Haynes2019}
{Haynes} C.~J.,  {Kobayashi} C.,  2019, \mn@doi [\mnras]
  {10.1093/mnras/sty3389}, \href
  {https://ui.adsabs.harvard.edu/abs/2019MNRAS.483.5123H} {483, 5123}

\bibitem[\protect\citeauthoryear{{Haywood}, {Di Matteo}, {Lehnert}, {Snaith},
  {Khoperskov}  \& {G{\'o}mez}}{{Haywood} et~al.}{2018}]{Haywood2018}
{Haywood} M.,  {Di Matteo} P.,  {Lehnert} M.~D.,  {Snaith} O.,  {Khoperskov}
  S.,   {G{\'o}mez} A.,  2018, \mn@doi [\apj] {10.3847/1538-4357/aad235}, \href
  {https://ui.adsabs.harvard.edu/abs/2018ApJ...863..113H} {863, 113}

\bibitem[\protect\citeauthoryear{{Helmi}, {White}, {de Zeeuw}  \&
  {Zhao}}{{Helmi} et~al.}{1999}]{helmi1999}
{Helmi} A.,  {White} S. D.~M.,  {de Zeeuw} P.~T.,   {Zhao} H.,  1999, \mn@doi
  [\nat] {10.1038/46980}, \href
  {https://ui.adsabs.harvard.edu/abs/1999Natur.402...53H} {402, 53}

\bibitem[\protect\citeauthoryear{{Helmi}, {Babusiaux}, {Koppelman}, {Massari},
  {Veljanoski}  \& {Brown}}{{Helmi} et~al.}{2018}]{helmi2018}
{Helmi} A.,  {Babusiaux} C.,  {Koppelman} H.~H.,  {Massari} D.,  {Veljanoski}
  J.,   {Brown} A. G.~A.,  2018, \mn@doi [\nat] {10.1038/s41586-018-0625-x},
  \href {https://ui.adsabs.harvard.edu/abs/2018Natur.563...85H} {563, 85}

\bibitem[\protect\citeauthoryear{{Hill} et~al.,}{{Hill}
  et~al.}{2019}]{Hill2019}
{Hill} V.,  et~al., 2019, \mn@doi [\aap] {10.1051/0004-6361/201833950}, \href
  {https://ui.adsabs.harvard.edu/abs/2019A&A...626A..15H} {626, A15}

\bibitem[\protect\citeauthoryear{{Horta} et~al.,}{{Horta}
  et~al.}{2023}]{Horta2023haloSubs}
{Horta} D.,  et~al., 2023, \mn@doi [\mnras] {10.1093/mnras/stac3179}, \href
  {https://ui.adsabs.harvard.edu/abs/2023MNRAS.520.5671H} {520, 5671}

\bibitem[\protect\citeauthoryear{{Ibata}, {Gilmore}  \& {Irwin}}{{Ibata}
  et~al.}{1994}]{Ibata1994}
{Ibata} R.~A.,  {Gilmore} G.,   {Irwin} M.~J.,  1994, \mn@doi [\nat]
  {10.1038/370194a0}, \href
  {https://ui.adsabs.harvard.edu/abs/1994Natur.370..194I} {370, 194}

\bibitem[\protect\citeauthoryear{{Jablonka} et~al.,}{{Jablonka}
  et~al.}{2015}]{jablonka2015scl}
{Jablonka} P.,  et~al., 2015, \mn@doi [\aap] {10.1051/0004-6361/201525661},
  \href {https://ui.adsabs.harvard.edu/abs/2015A&A...583A..67J} {583, A67}

\bibitem[\protect\citeauthoryear{{Jacobson} et~al.,}{{Jacobson}
  et~al.}{2015}]{Jacobson2015metalpoor}
{Jacobson} H.~R.,  et~al., 2015, \mn@doi [\apj] {10.1088/0004-637X/807/2/171},
  \href {https://ui.adsabs.harvard.edu/abs/2015ApJ...807..171J} {807, 171}

\bibitem[\protect\citeauthoryear{{Ji}, {Frebel}, {Chiti}  \& {Simon}}{{Ji}
  et~al.}{2016a}]{ji2016a}
{Ji} A.~P.,  {Frebel} A.,  {Chiti} A.,   {Simon} J.~D.,  2016a, \mn@doi [\nat]
  {10.1038/nature17425}, \href
  {https://ui.adsabs.harvard.edu/abs/2016Natur.531..610J} {531, 610}

\bibitem[\protect\citeauthoryear{{Ji}, {Frebel}, {Simon}  \& {Chiti}}{{Ji}
  et~al.}{2016b}]{Ji2016b}
{Ji} A.~P.,  {Frebel} A.,  {Simon} J.~D.,   {Chiti} A.,  2016b, \mn@doi [\apj]
  {10.3847/0004-637X/830/2/93}, \href
  {https://ui.adsabs.harvard.edu/abs/2016ApJ...830...93J} {830, 93}

\bibitem[\protect\citeauthoryear{{Ji} et~al.,}{{Ji}
  et~al.}{2020a}]{Ji2020streams}
{Ji} A.~P.,  et~al., 2020a, \mn@doi [\aj] {10.3847/1538-3881/abacb6}, \href
  {https://ui.adsabs.harvard.edu/abs/2020AJ....160..181J} {160, 181}

\bibitem[\protect\citeauthoryear{{Ji} et~al.,}{{Ji}
  et~al.}{2020b}]{Ji2020MagLiteS}
{Ji} A.~P.,  et~al., 2020b, \mn@doi [\apj] {10.3847/1538-4357/ab6213}, \href
  {https://ui.adsabs.harvard.edu/abs/2020ApJ...889...27J} {889, 27}

\bibitem[\protect\citeauthoryear{{Ji} et~al.,}{{Ji}
  et~al.}{2023a}]{AlexJi2023ret2}
{Ji} A.~P.,  et~al., 2023a, \mn@doi [\aj] {10.3847/1538-3881/acad84}, \href
  {https://ui.adsabs.harvard.edu/abs/2023AJ....165..100J} {165, 100}

\bibitem[\protect\citeauthoryear{{Ji}, {Naidu}, {Brauer}, {Ting}  \&
  {Simon}}{{Ji} et~al.}{2023b}]{AlexJi2023typhon}
{Ji} A.~P.,  {Naidu} R.~P.,  {Brauer} K.,  {Ting} Y.-S.,   {Simon} J.~D.,
  2023b, \mn@doi [\mnras] {10.1093/mnras/stac2757}, \href
  {https://ui.adsabs.harvard.edu/abs/2023MNRAS.519.4467J} {519, 4467}

\bibitem[\protect\citeauthoryear{{Johnson} et~al.,}{{Johnson}
  et~al.}{2020}]{Johnson2020sgr}
{Johnson} B.~D.,  et~al., 2020, \mn@doi [\apj] {10.3847/1538-4357/abab08},
  \href {https://ui.adsabs.harvard.edu/abs/2020ApJ...900..103J} {900, 103}

\bibitem[\protect\citeauthoryear{{Johnson} et~al.,}{{Johnson}
  et~al.}{2022}]{Johnson2022wukong}
{Johnson} J.~W.,  et~al., 2022, \mn@doi [arXiv e-prints]
  {10.48550/arXiv.2210.01816}, \href
  {https://ui.adsabs.harvard.edu/abs/2022arXiv221001816J} {p. arXiv:2210.01816}

\bibitem[\protect\citeauthoryear{{Johnston}}{{Johnston}}{1998}]{Johnston1998}
{Johnston} K.~V.,  1998, \mn@doi [\apj] {10.1086/305273}, \href
  {https://ui.adsabs.harvard.edu/abs/1998ApJ...495..297J} {495, 297}

\bibitem[\protect\citeauthoryear{{Kauffmann}, {White}  \&
  {Guiderdoni}}{{Kauffmann} et~al.}{1993}]{Kauffmann1993}
{Kauffmann} G.,  {White} S.~D.~M.,   {Guiderdoni} B.,  1993, \mn@doi [\mnras]
  {10.1093/mnras/264.1.201}, \href
  {https://ui.adsabs.harvard.edu/abs/1993MNRAS.264..201K} {264, 201}

\bibitem[\protect\citeauthoryear{{Kelson}}{{Kelson}}{2003}]{carpy}
{Kelson} D.~D.,  2003, \mn@doi [\pasp] {10.1086/375502}, \href
  {https://ui.adsabs.harvard.edu/abs/2003PASP..115..688K} {115, 688}

\bibitem[\protect\citeauthoryear{{Kilpatrick} et~al.,}{{Kilpatrick}
  et~al.}{2017}]{Kilpatrick2017}
{Kilpatrick} C.~D.,  et~al., 2017, \mn@doi [Science] {10.1126/science.aaq0073},
  \href {https://ui.adsabs.harvard.edu/abs/2017Sci...358.1583K} {358, 1583}

\bibitem[\protect\citeauthoryear{{Kirby}, {Guhathakurta}, {Bolte}, {Sneden}  \&
  {Geha}}{{Kirby} et~al.}{2009}]{Kirby2009sculptor}
{Kirby} E.~N.,  {Guhathakurta} P.,  {Bolte} M.,  {Sneden} C.,   {Geha} M.~C.,
  2009, \mn@doi [\apj] {10.1088/0004-637X/705/1/328}, \href
  {https://ui.adsabs.harvard.edu/abs/2009ApJ...705..328K} {705, 328}

\bibitem[\protect\citeauthoryear{{Kirby}, {Cohen}, {Smith}, {Majewski}, {Sohn}
  \& {Guhathakurta}}{{Kirby} et~al.}{2011}]{Kirby2011alphas}
{Kirby} E.~N.,  {Cohen} J.~G.,  {Smith} G.~H.,  {Majewski} S.~R.,  {Sohn}
  S.~T.,   {Guhathakurta} P.,  2011, \mn@doi [\apj]
  {10.1088/0004-637X/727/2/79}, \href
  {https://ui.adsabs.harvard.edu/abs/2011ApJ...727...79K} {727, 79}

\bibitem[\protect\citeauthoryear{{Kobayashi}, {Karakas}  \&
  {Lugaro}}{{Kobayashi} et~al.}{2020}]{Kobayashi2020}
{Kobayashi} C.,  {Karakas} A.~I.,   {Lugaro} M.,  2020, \mn@doi [\apj]
  {10.3847/1538-4357/abae65}, \href
  {https://ui.adsabs.harvard.edu/abs/2020ApJ...900..179K} {900, 179}

\bibitem[\protect\citeauthoryear{{Koppelman}, {Helmi}, {Massari}, {Roelenga}
  \& {Bastian}}{{Koppelman} et~al.}{2019a}]{koppelmanHelmi}
{Koppelman} H.~H.,  {Helmi} A.,  {Massari} D.,  {Roelenga} S.,   {Bastian} U.,
  2019a, \mn@doi [\aap] {10.1051/0004-6361/201834769}, \href
  {https://ui.adsabs.harvard.edu/abs/2019A&A...625A...5K} {625, A5}

\bibitem[\protect\citeauthoryear{{Koppelman}, {Helmi}, {Massari},
  {Price-Whelan}  \& {Starkenburg}}{{Koppelman} et~al.}{2019b}]{koppelman2019}
{Koppelman} H.~H.,  {Helmi} A.,  {Massari} D.,  {Price-Whelan} A.~M.,
  {Starkenburg} T.~K.,  2019b, \mn@doi [\aap] {10.1051/0004-6361/201936738},
  \href {https://ui.adsabs.harvard.edu/abs/2019A&A...631L...9K} {631, L9}

\bibitem[\protect\citeauthoryear{{Koppelman}, {Bos}  \& {Helmi}}{{Koppelman}
  et~al.}{2020}]{Koppelman2020MassiveMerger}
{Koppelman} H.~H.,  {Bos} R. O.~Y.,   {Helmi} A.,  2020, \mn@doi [\aap]
  {10.1051/0004-6361/202038652}, \href
  {https://ui.adsabs.harvard.edu/abs/2020A&A...642L..18K} {642, L18}

\bibitem[\protect\citeauthoryear{{Kruijssen}, {Pfeffer}, {Reina-Campos},
  {Crain}  \& {Bastian}}{{Kruijssen} et~al.}{2019}]{Kruijssen2019milkyway}
{Kruijssen} J.~M.~D.,  {Pfeffer} J.~L.,  {Reina-Campos} M.,  {Crain} R.~A.,
  {Bastian} N.,  2019, \mn@doi [\mnras] {10.1093/mnras/sty1609}, \href
  {https://ui.adsabs.harvard.edu/abs/2019MNRAS.486.3180K} {486, 3180}

\bibitem[\protect\citeauthoryear{{Kruijssen} et~al.,}{{Kruijssen}
  et~al.}{2020}]{Kruijssen2020kraken}
{Kruijssen} J.~M.~D.,  et~al., 2020, \mn@doi [\mnras] {10.1093/mnras/staa2452},
  \href {https://ui.adsabs.harvard.edu/abs/2020MNRAS.498.2472K} {498, 2472}

\bibitem[\protect\citeauthoryear{{Larsen}, {Brodie}, {Grundahl}  \&
  {Strader}}{{Larsen} et~al.}{2014}]{Larsen2014fornax}
{Larsen} S.~S.,  {Brodie} J.~P.,  {Grundahl} F.,   {Strader} J.,  2014, \mn@doi
  [\apj] {10.1088/0004-637X/797/1/15}, \href
  {https://ui.adsabs.harvard.edu/abs/2014ApJ...797...15L} {797, 15}

\bibitem[\protect\citeauthoryear{{Lee} et~al.,}{{Lee} et~al.}{2013}]{lee2013}
{Lee} Y.~S.,  et~al., 2013, \mn@doi [\aj] {10.1088/0004-6256/146/5/132}, \href
  {https://ui.adsabs.harvard.edu/abs/2013AJ....146..132L} {146, 132}

\bibitem[\protect\citeauthoryear{{Letarte} et~al.,}{{Letarte}
  et~al.}{2010}]{Letarte2010}
{Letarte} B.,  et~al., 2010, \mn@doi [\aap] {10.1051/0004-6361/200913413},
  \href {https://ui.adsabs.harvard.edu/abs/2010A&A...523A..17L} {523, A17}

\bibitem[\protect\citeauthoryear{{Li} et~al.,}{{Li}
  et~al.}{2022}]{Li2022lamost}
{Li} H.,  et~al., 2022, \mn@doi [\apj] {10.3847/1538-4357/ac6514}, \href
  {https://ui.adsabs.harvard.edu/abs/2022ApJ...931..147L} {931, 147}

\bibitem[\protect\citeauthoryear{{Limberg} et~al.,}{{Limberg}
  et~al.}{2021}]{Limberg2021hstr}
{Limberg} G.,  et~al., 2021, \mn@doi [\apjl] {10.3847/2041-8213/ac0056}, \href
  {https://ui.adsabs.harvard.edu/abs/2021ApJ...913L..28L} {913, L28}

\bibitem[\protect\citeauthoryear{{Limberg}, {Souza}, {P{\'e}rez-Villegas},
  {Rossi}, {Perottoni}  \& {Santucci}}{{Limberg} et~al.}{2022}]{Limberg2022gse}
{Limberg} G.,  {Souza} S.~O.,  {P{\'e}rez-Villegas} A.,  {Rossi} S.,
  {Perottoni} H.~D.,   {Santucci} R.~M.,  2022, \mn@doi [\apj]
  {10.3847/1538-4357/ac8159}, \href
  {https://ui.adsabs.harvard.edu/abs/2022ApJ...935..109L} {935, 109}

\bibitem[\protect\citeauthoryear{{Limberg} et~al.,}{{Limberg}
  et~al.}{2023}]{Limberg2023sgr}
{Limberg} G.,  et~al., 2023, \mn@doi [\apj] {10.3847/1538-4357/acb694}, \href
  {https://ui.adsabs.harvard.edu/abs/2023ApJ...946...66L} {946, 66}

\bibitem[\protect\citeauthoryear{{Lind}, {Asplund}, {Barklem}  \&
  {Belyaev}}{{Lind} et~al.}{2011}]{Lind2011nlte_forNa}
{Lind} K.,  {Asplund} M.,  {Barklem} P.~S.,   {Belyaev} A.~K.,  2011, \mn@doi
  [\aap] {10.1051/0004-6361/201016095}, \href
  {https://ui.adsabs.harvard.edu/abs/2011A&A...528A.103L} {528, A103}

\bibitem[\protect\citeauthoryear{{Lucatello}, {Tsangarides}, {Beers},
  {Carretta}, {Gratton}  \& {Ryan}}{{Lucatello} et~al.}{2005}]{Lucatello2005}
{Lucatello} S.,  {Tsangarides} S.,  {Beers} T.~C.,  {Carretta} E.,  {Gratton}
  R.~G.,   {Ryan} S.~G.,  2005, \mn@doi [\apj] {10.1086/428104}, \href
  {https://ui.adsabs.harvard.edu/abs/2005ApJ...625..825L} {625, 825}

\bibitem[\protect\citeauthoryear{{Lucatello}, {Beers}, {Christlieb}, {Barklem},
  {Rossi}, {Marsteller}, {Sivarani}  \& {Lee}}{{Lucatello}
  et~al.}{2006}]{lucatello2006}
{Lucatello} S.,  {Beers} T.~C.,  {Christlieb} N.,  {Barklem} P.~S.,  {Rossi}
  S.,  {Marsteller} B.,  {Sivarani} T.,   {Lee} Y.~S.,  2006, \mn@doi [\apjl]
  {10.1086/509780}, \href
  {https://ui.adsabs.harvard.edu/abs/2006ApJ...652L..37L} {652, L37}

\bibitem[\protect\citeauthoryear{{Lucey} et~al.,}{{Lucey}
  et~al.}{2023}]{Lucey2023cemp}
{Lucey} M.,  et~al., 2023, \mn@doi [\mnras] {10.1093/mnras/stad1675}, \href
  {https://ui.adsabs.harvard.edu/abs/2023MNRAS.523.4049L} {523, 4049}

\bibitem[\protect\citeauthoryear{{Majewski}, {Skrutskie}, {Weinberg}  \&
  {Ostheimer}}{{Majewski} et~al.}{2003}]{Majewski2003}
{Majewski} S.~R.,  {Skrutskie} M.~F.,  {Weinberg} M.~D.,   {Ostheimer} J.~C.,
  2003, \mn@doi [\apj] {10.1086/379504}, \href
  {https://ui.adsabs.harvard.edu/abs/2003ApJ...599.1082M} {599, 1082}

\bibitem[\protect\citeauthoryear{{Majewski} et~al.,}{{Majewski}
  et~al.}{2017}]{apogee2017}
{Majewski} S.~R.,  et~al., 2017, \mn@doi [\aj] {10.3847/1538-3881/aa784d},
  \href {https://ui.adsabs.harvard.edu/abs/2017AJ....154...94M} {154, 94}

\bibitem[\protect\citeauthoryear{{Malhan}, {Yuan}, {Ibata}, {Arentsen},
  {Bellazzini}  \& {Martin}}{{Malhan} et~al.}{2021}]{Malhan2021lms1}
{Malhan} K.,  {Yuan} Z.,  {Ibata} R.~A.,  {Arentsen} A.,  {Bellazzini} M.,
  {Martin} N.~F.,  2021, \mn@doi [\apj] {10.3847/1538-4357/ac1675}, \href
  {https://ui.adsabs.harvard.edu/abs/2021ApJ...920...51M} {920, 51}

\bibitem[\protect\citeauthoryear{{Malhan} et~al.,}{{Malhan}
  et~al.}{2022}]{Malhan2022atlas}
{Malhan} K.,  et~al., 2022, \mn@doi [\apj] {10.3847/1538-4357/ac4d2a}, \href
  {https://ui.adsabs.harvard.edu/abs/2022ApJ...926..107M} {926, 107}

\bibitem[\protect\citeauthoryear{{Maoz} \& {Mannucci}}{{Maoz} \&
  {Mannucci}}{2012}]{Maoz2012dtdSNIa}
{Maoz} D.,  {Mannucci} F.,  2012, \mn@doi [\pasa] {10.1071/AS11052}, \href
  {https://ui.adsabs.harvard.edu/abs/2012PASA...29..447M} {29, 447}

\bibitem[\protect\citeauthoryear{{Marino}, {Villanova}, {Piotto}, {Milone},
  {Momany}, {Bedin}  \& {Medling}}{{Marino} et~al.}{2008}]{Marino2008m4}
{Marino} A.~F.,  {Villanova} S.,  {Piotto} G.,  {Milone} A.~P.,  {Momany} Y.,
  {Bedin} L.~R.,   {Medling} A.~M.,  2008, \mn@doi [\aap]
  {10.1051/0004-6361:200810389}, \href
  {https://ui.adsabs.harvard.edu/abs/2008A&A...490..625M} {490, 625}

\bibitem[\protect\citeauthoryear{{Martin} et~al.,}{{Martin}
  et~al.}{2022}]{Martin2022c19}
{Martin} N.~F.,  et~al., 2022, \mn@doi [\nat] {10.1038/s41586-021-04162-2},
  \href {https://ui.adsabs.harvard.edu/abs/2022Natur.601...45M} {601, 45}

\bibitem[\protect\citeauthoryear{{Mateu}}{{Mateu}}{2023}]{Mateu2023galstreams}
{Mateu} C.,  2023, \mn@doi [\mnras] {10.1093/mnras/stad321}, \href
  {https://ui.adsabs.harvard.edu/abs/2023MNRAS.520.5225M} {520, 5225}

\bibitem[\protect\citeauthoryear{{Matsuno}, {Aoki}  \& {Suda}}{{Matsuno}
  et~al.}{2019}]{matsuno2019}
{Matsuno} T.,  {Aoki} W.,   {Suda} T.,  2019, \mn@doi [\apjl]
  {10.3847/2041-8213/ab0ec0}, \href
  {https://ui.adsabs.harvard.edu/abs/2019ApJ...874L..35M} {874, L35}

\bibitem[\protect\citeauthoryear{{Matsuno}, {Hirai}, {Tarumi}, {Hotokezaka},
  {Tanaka}  \& {Helmi}}{{Matsuno} et~al.}{2021}]{Matsuno2021gseRprocess}
{Matsuno} T.,  {Hirai} Y.,  {Tarumi} Y.,  {Hotokezaka} K.,  {Tanaka} M.,
  {Helmi} A.,  2021, \mn@doi [\aap] {10.1051/0004-6361/202040227}, \href
  {https://ui.adsabs.harvard.edu/abs/2021A&A...650A.110M} {650, A110}

\bibitem[\protect\citeauthoryear{{Matsuno}, {Koppelman}, {Helmi}, {Aoki},
  {Ishigaki}, {Suda}, {Yuan}  \& {Hattori}}{{Matsuno}
  et~al.}{2022a}]{Matsuno2022seq}
{Matsuno} T.,  {Koppelman} H.~H.,  {Helmi} A.,  {Aoki} W.,  {Ishigaki} M.~N.,
  {Suda} T.,  {Yuan} Z.,   {Hattori} K.,  2022a, \mn@doi [\aap]
  {10.1051/0004-6361/202142752}, \href
  {https://ui.adsabs.harvard.edu/abs/2022A&A...661A.103M} {661, A103}

\bibitem[\protect\citeauthoryear{{Matsuno} et~al.,}{{Matsuno}
  et~al.}{2022b}]{Matsuno2022hstr}
{Matsuno} T.,  et~al., 2022b, \mn@doi [\aap] {10.1051/0004-6361/202243609},
  \href {https://ui.adsabs.harvard.edu/abs/2022A&A...665A..46M} {665, A46}

\bibitem[\protect\citeauthoryear{{Matteucci}}{{Matteucci}}{2012}]{Matteucci2012book}
{Matteucci} F.,  2012, {Chemical Evolution of Galaxies}.
Berlin: Springer

\bibitem[\protect\citeauthoryear{{Matteucci} \& {Brocato}}{{Matteucci} \&
  {Brocato}}{1990}]{Matteucci1990}
{Matteucci} F.,  {Brocato} E.,  1990, \mn@doi [\apj] {10.1086/169508}, \href
  {https://ui.adsabs.harvard.edu/abs/1990ApJ...365..539M} {365, 539}

\bibitem[\protect\citeauthoryear{{Matteucci} \& {Greggio}}{{Matteucci} \&
  {Greggio}}{1986}]{Matteucci1986}
{Matteucci} F.,  {Greggio} L.,  1986, \aap, \href
  {https://ui.adsabs.harvard.edu/abs/1986A&A...154..279M} {154, 279}

\bibitem[\protect\citeauthoryear{{McConnachie}}{{McConnachie}}{2012}]{McConnachie2012catalog}
{McConnachie} A.~W.,  2012, \mn@doi [\aj] {10.1088/0004-6256/144/1/4}, \href
  {https://ui.adsabs.harvard.edu/abs/2012AJ....144....4M} {144, 4}

\bibitem[\protect\citeauthoryear{{McMillan}}{{McMillan}}{2017}]{mcmillan2017}
{McMillan} P.~J.,  2017, \mn@doi [\mnras] {10.1093/mnras/stw2759}, \href
  {https://ui.adsabs.harvard.edu/abs/2017MNRAS.465...76M} {465, 76}

\bibitem[\protect\citeauthoryear{{McWilliam}, {Wallerstein}  \&
  {Mottini}}{{McWilliam} et~al.}{2013}]{McWilliam2013params}
{McWilliam} A.,  {Wallerstein} G.,   {Mottini} M.,  2013, \mn@doi [\apj]
  {10.1088/0004-637X/778/2/149}, \href
  {https://ui.adsabs.harvard.edu/abs/2013ApJ...778..149M} {778, 149}

\bibitem[\protect\citeauthoryear{{Milone} \& {Marino}}{{Milone} \&
  {Marino}}{2022}]{Milone2022mps}
{Milone} A.~P.,  {Marino} A.~F.,  2022, \mn@doi [Universe]
  {10.3390/universe8070359}, \href
  {https://ui.adsabs.harvard.edu/abs/2022Univ....8..359M} {8, 359}

\bibitem[\protect\citeauthoryear{{Milone} et~al.,}{{Milone}
  et~al.}{2017}]{milone2017}
{Milone} A.~P.,  et~al., 2017, \mn@doi [\mnras] {10.1093/mnras/stw2531}, \href
  {https://ui.adsabs.harvard.edu/abs/2017MNRAS.464.3636M} {464, 3636}

\bibitem[\protect\citeauthoryear{{Monty}, {Venn}, {Lane}, {Lokhorst}  \&
  {Yong}}{{Monty} et~al.}{2020}]{Monty2020}
{Monty} S.,  {Venn} K.~A.,  {Lane} J. M.~M.,  {Lokhorst} D.,   {Yong} D.,
  2020, \mn@doi [\mnras] {10.1093/mnras/staa1995}, \href
  {https://ui.adsabs.harvard.edu/abs/2020MNRAS.497.1236M} {497, 1236}

\bibitem[\protect\citeauthoryear{{Morinaga}, {Ishiyama}, {Kirihara}  \&
  {Kinjo}}{{Morinaga} et~al.}{2019}]{Morinaga2019_MWhalos}
{Morinaga} Y.,  {Ishiyama} T.,  {Kirihara} T.,   {Kinjo} K.,  2019, \mn@doi
  [\mnras] {10.1093/mnras/stz1373}, \href
  {https://ui.adsabs.harvard.edu/abs/2019MNRAS.487.2718M} {487, 2718}

\bibitem[\protect\citeauthoryear{{Mucciarelli}, {Origlia}, {Ferraro}  \&
  {Pancino}}{{Mucciarelli} et~al.}{2009}]{Mucciarelli2009lmc}
{Mucciarelli} A.,  {Origlia} L.,  {Ferraro} F.~R.,   {Pancino} E.,  2009,
  \mn@doi [\apjl] {10.1088/0004-637X/695/2/L134}, \href
  {https://ui.adsabs.harvard.edu/abs/2009ApJ...695L.134M} {695, L134}

\bibitem[\protect\citeauthoryear{{Myeong}, {Evans}, {Belokurov}, {Amorisco}  \&
  {Koposov}}{{Myeong} et~al.}{2018}]{myeongStreamsAndClumps}
{Myeong} G.~C.,  {Evans} N.~W.,  {Belokurov} V.,  {Amorisco} N.~C.,   {Koposov}
  S.~E.,  2018, \mn@doi [\mnras] {10.1093/mnras/stx3262}, \href
  {https://ui.adsabs.harvard.edu/abs/2018MNRAS.475.1537M} {475, 1537}

\bibitem[\protect\citeauthoryear{{Myeong}, {Vasiliev}, {Iorio}, {Evans}  \&
  {Belokurov}}{{Myeong} et~al.}{2019}]{myeongSequoia}
{Myeong} G.~C.,  {Vasiliev} E.,  {Iorio} G.,  {Evans} N.~W.,   {Belokurov} V.,
  2019, \mn@doi [\mnras] {10.1093/mnras/stz1770}, \href
  {https://ui.adsabs.harvard.edu/abs/2019MNRAS.488.1235M} {488, 1235}

\bibitem[\protect\citeauthoryear{{Naidu}, {Conroy}, {Bonaca}, {Johnson},
  {Ting}, {Caldwell}, {Zaritsky}  \& {Cargile}}{{Naidu}
  et~al.}{2020}]{naidu2020}
{Naidu} R.~P.,  {Conroy} C.,  {Bonaca} A.,  {Johnson} B.~D.,  {Ting} Y.-S.,
  {Caldwell} N.,  {Zaritsky} D.,   {Cargile} P.~A.,  2020, \mn@doi [\apj]
  {10.3847/1538-4357/abaef4}, \href
  {https://ui.adsabs.harvard.edu/abs/2020ApJ...901...48N} {901, 48}

\bibitem[\protect\citeauthoryear{{Naidu} et~al.,}{{Naidu}
  et~al.}{2021}]{Naidu2021simulations}
{Naidu} R.~P.,  et~al., 2021, \mn@doi [\apj] {10.3847/1538-4357/ac2d2d}, \href
  {https://ui.adsabs.harvard.edu/abs/2021ApJ...923...92N} {923, 92}

\bibitem[\protect\citeauthoryear{{Naidu} et~al.,}{{Naidu}
  et~al.}{2022}]{Naidu2022mzr}
{Naidu} R.~P.,  et~al., 2022, arXiv e-prints, \href
  {https://ui.adsabs.harvard.edu/abs/2022arXiv220409057N} {p. arXiv:2204.09057}

\bibitem[\protect\citeauthoryear{{Nidever} et~al.,}{{Nidever}
  et~al.}{2020}]{Nidever2020}
{Nidever} D.~L.,  et~al., 2020, \mn@doi [\apj] {10.3847/1538-4357/ab7305},
  \href {https://ui.adsabs.harvard.edu/abs/2020ApJ...895...88N} {895, 88}

\bibitem[\protect\citeauthoryear{{Nissen} \& {Schuster}}{{Nissen} \&
  {Schuster}}{2010}]{NissenSchuster2010}
{Nissen} P.~E.,  {Schuster} W.~J.,  2010, \mn@doi [\aap]
  {10.1051/0004-6361/200913877}, \href
  {https://ui.adsabs.harvard.edu/abs/2010A&A...511L..10N} {511, L10}

\bibitem[\protect\citeauthoryear{{Nissen} \& {Schuster}}{{Nissen} \&
  {Schuster}}{2011}]{NissenSchuster2011}
{Nissen} P.~E.,  {Schuster} W.~J.,  2011, \mn@doi [\aap]
  {10.1051/0004-6361/201116619}, \href
  {https://ui.adsabs.harvard.edu/abs/2011A&A...530A..15N} {530, A15}

\bibitem[\protect\citeauthoryear{{Nomoto}, {Kobayashi}  \& {Tominaga}}{{Nomoto}
  et~al.}{2013}]{Nomoto2013}
{Nomoto} K.,  {Kobayashi} C.,   {Tominaga} N.,  2013, \mn@doi [\araa]
  {10.1146/annurev-astro-082812-140956}, \href
  {https://ui.adsabs.harvard.edu/abs/2013ARA&A..51..457N} {51, 457}

\bibitem[\protect\citeauthoryear{{Norris}, {Ryan}  \& {Beers}}{{Norris}
  et~al.}{1997}]{norris1997}
{Norris} J.~E.,  {Ryan} S.~G.,   {Beers} T.~C.,  1997, \mn@doi [\apj]
  {10.1086/304695}, \href
  {https://ui.adsabs.harvard.edu/abs/1997ApJ...488..350N} {488, 350}

\bibitem[\protect\citeauthoryear{{Pe{\~n}arrubia} \&
  {Petersen}}{{Pe{\~n}arrubia} \& {Petersen}}{2021}]{Penarrubia2021sgr}
{Pe{\~n}arrubia} J.,  {Petersen} M.~S.,  2021, \mn@doi [\mnras]
  {10.1093/mnrasl/slab090}, \href
  {https://ui.adsabs.harvard.edu/abs/2021MNRAS.508L..26P} {508, L26}

\bibitem[\protect\citeauthoryear{{Pillepich}, {Madau}  \& {Mayer}}{{Pillepich}
  et~al.}{2015}]{Pillepich2015halos}
{Pillepich} A.,  {Madau} P.,   {Mayer} L.,  2015, \mn@doi [\apj]
  {10.1088/0004-637X/799/2/184}, \href
  {https://ui.adsabs.harvard.edu/abs/2015ApJ...799..184P} {799, 184}

\bibitem[\protect\citeauthoryear{{Piotto} et~al.,}{{Piotto}
  et~al.}{2015}]{piotto2015}
{Piotto} G.,  et~al., 2015, \mn@doi [\aj] {10.1088/0004-6256/149/3/91}, \href
  {https://ui.adsabs.harvard.edu/abs/2015AJ....149...91P} {149, 91}

\bibitem[\protect\citeauthoryear{{Placco}, {Frebel}, {Beers}  \&
  {Stancliffe}}{{Placco} et~al.}{2014}]{placco2014Carbon}
{Placco} V.~M.,  {Frebel} A.,  {Beers} T.~C.,   {Stancliffe} R.~J.,  2014,
  \mn@doi [\apj] {10.1088/0004-637X/797/1/21}, \href
  {https://ui.adsabs.harvard.edu/abs/2014ApJ...797...21P} {797, 21}

\bibitem[\protect\citeauthoryear{{Placco} et~al.,}{{Placco}
  et~al.}{2018}]{placco2018}
{Placco} V.~M.,  et~al., 2018, \mn@doi [\aj] {10.3847/1538-3881/aac20c}, \href
  {https://ui.adsabs.harvard.edu/abs/2018AJ....155..256P} {155, 256}

\bibitem[\protect\citeauthoryear{{Placco}, {Sneden}, {Roederer}, {Lawler}, {Den
  Hartog}, {Hejazi}, {Maas}  \& {Bernath}}{{Placco}
  et~al.}{2021}]{Placco2021linemake}
{Placco} V.~M.,  {Sneden} C.,  {Roederer} I.~U.,  {Lawler} J.~E.,  {Den Hartog}
  E.~A.,  {Hejazi} N.,  {Maas} Z.,   {Bernath} P.,  2021, \mn@doi [Research
  Notes of the American Astronomical Society] {10.3847/2515-5172/abf651}, \href
  {https://ui.adsabs.harvard.edu/abs/2021RNAAS...5...92P} {5, 92}

\bibitem[\protect\citeauthoryear{{Planck Collaboration} et~al.,}{{Planck
  Collaboration} et~al.}{2020}]{PlanckCollab2020}
{Planck Collaboration} et~al., 2020, \mn@doi [\aap]
  {10.1051/0004-6361/201833910}, \href
  {https://ui.adsabs.harvard.edu/abs/2020A&A...641A...6P} {641, A6}

\bibitem[\protect\citeauthoryear{{Queiroz} et~al.,}{{Queiroz}
  et~al.}{2020}]{Queiroz2020}
{Queiroz} A.~B.~A.,  et~al., 2020, \mn@doi [\aap]
  {10.1051/0004-6361/201937364}, \href
  {https://ui.adsabs.harvard.edu/abs/2020A&A...638A..76Q} {638, A76}

\bibitem[\protect\citeauthoryear{{Queiroz} et~al.,}{{Queiroz}
  et~al.}{2023}]{Queiroz2023starhorse}
{Queiroz} A.~B.~A.,  et~al., 2023, \mn@doi [\aap]
  {10.1051/0004-6361/202245399}, \href
  {https://ui.adsabs.harvard.edu/abs/2023A&A...673A.155Q} {673, A155}

\bibitem[\protect\citeauthoryear{{Reichert}, {Hansen}, {Hanke},
  {Sk{\'u}lad{\'o}ttir}, {Arcones}  \& {Grebel}}{{Reichert}
  et~al.}{2020}]{Reichert2020alphaKnees}
{Reichert} M.,  {Hansen} C.~J.,  {Hanke} M.,  {Sk{\'u}lad{\'o}ttir} {\'A}.,
  {Arcones} A.,   {Grebel} E.~K.,  2020, \mn@doi [\aap]
  {10.1051/0004-6361/201936930}, \href
  {https://ui.adsabs.harvard.edu/abs/2020A&A...641A.127R} {641, A127}

\bibitem[\protect\citeauthoryear{{Rodr{\'\i}guez-Puebla}, {Primack},
  {Avila-Reese}  \& {Faber}}{{Rodr{\'\i}guez-Puebla}
  et~al.}{2017}]{Rodriguez-Puebla2017}
{Rodr{\'\i}guez-Puebla} A.,  {Primack} J.~R.,  {Avila-Reese} V.,   {Faber}
  S.~M.,  2017, \mn@doi [\mnras] {10.1093/mnras/stx1172}, \href
  {https://ui.adsabs.harvard.edu/abs/2017MNRAS.470..651R} {470, 651}

\bibitem[\protect\citeauthoryear{{Roederer}, {Sneden}, {Thompson}, {Preston}
  \& {Shectman}}{{Roederer} et~al.}{2010}]{Roederer2010}
{Roederer} I.~U.,  {Sneden} C.,  {Thompson} I.~B.,  {Preston} G.~W.,
  {Shectman} S.~A.,  2010, \mn@doi [\apj] {10.1088/0004-637X/711/2/573}, \href
  {https://ui.adsabs.harvard.edu/abs/2010ApJ...711..573R} {711, 573}

\bibitem[\protect\citeauthoryear{{Roederer}, {Preston}, {Thompson}, {Shectman},
  {Sneden}, {Burley}  \& {Kelson}}{{Roederer} et~al.}{2014}]{roederer2014}
{Roederer} I.~U.,  {Preston} G.~W.,  {Thompson} I.~B.,  {Shectman} S.~A.,
  {Sneden} C.,  {Burley} G.~S.,   {Kelson} D.~D.,  2014, \mn@doi [\aj]
  {10.1088/0004-6256/147/6/136}, \href
  {https://ui.adsabs.harvard.edu/abs/2014AJ....147..136R} {147, 136}

\bibitem[\protect\citeauthoryear{{Rossi}, {Beers}  \& {Sneden}}{{Rossi}
  et~al.}{1999}]{rossi1999}
{Rossi} S.,  {Beers} T.~C.,   {Sneden} C.,  1999, in {Gibson} B.~K.,  {Axelrod}
  R.~S.,   {Putman} M.~E.,  eds,  Astronomical Society of the Pacific
  Conference Series Vol. 165, The Third Stromlo Symposium: The Galactic Halo.
  p.~264

\bibitem[\protect\citeauthoryear{{Rossi}, {Beers}, {Sneden}, {Sevastyanenko},
  {Rhee}  \& {Marsteller}}{{Rossi} et~al.}{2005}]{rossi2005}
{Rossi} S.,  {Beers} T.~C.,  {Sneden} C.,  {Sevastyanenko} T.,  {Rhee} J.,
  {Marsteller} B.,  2005, \mn@doi [\aj] {10.1086/497164}, \href
  {https://ui.adsabs.harvard.edu/abs/2005AJ....130.2804R} {130, 2804}

\bibitem[\protect\citeauthoryear{{Sch{\"o}nrich}, {Binney}  \&
  {Dehnen}}{{Sch{\"o}nrich} et~al.}{2010}]{schon2010}
{Sch{\"o}nrich} R.,  {Binney} J.,   {Dehnen} W.,  2010, \mn@doi [\mnras]
  {10.1111/j.1365-2966.2010.16253.x}, \href
  {https://ui.adsabs.harvard.edu/abs/2010MNRAS.403.1829S} {403, 1829}

\bibitem[\protect\citeauthoryear{{Searle} \& {Zinn}}{{Searle} \&
  {Zinn}}{1978}]{sz1978}
{Searle} L.,  {Zinn} R.,  1978, \mn@doi [\apj] {10.1086/156499}, \href
  {https://ui.adsabs.harvard.edu/abs/1978ApJ...225..357S} {225, 357}

\bibitem[\protect\citeauthoryear{{Shappee} et~al.,}{{Shappee}
  et~al.}{2017}]{shappee2017}
{Shappee} B.~J.,  et~al., 2017, \mn@doi [Science] {10.1126/science.aaq0186},
  \href {https://ui.adsabs.harvard.edu/abs/2017Sci...358.1574S} {358, 1574}

\bibitem[\protect\citeauthoryear{{Sharpe}, {Naidu}  \& {Conroy}}{{Sharpe}
  et~al.}{2022}]{Sharpe2022halo}
{Sharpe} K.,  {Naidu} R.~P.,   {Conroy} C.,  2022, arXiv e-prints, \href
  {https://ui.adsabs.harvard.edu/abs/2022arXiv221104562S} {p. arXiv:2211.04562}

\bibitem[\protect\citeauthoryear{{Shetrone}, {C{\^o}t{\'e}}  \&
  {Sargent}}{{Shetrone} et~al.}{2001}]{Shetrone2001}
{Shetrone} M.~D.,  {C{\^o}t{\'e}} P.,   {Sargent} W.~L.~W.,  2001, \mn@doi
  [\apj] {10.1086/319022}, \href
  {https://ui.adsabs.harvard.edu/abs/2001ApJ...548..592S} {548, 592}

\bibitem[\protect\citeauthoryear{{Shetrone}, {Venn}, {Tolstoy}, {Primas},
  {Hill}  \& {Kaufer}}{{Shetrone} et~al.}{2003}]{Shetrone2003}
{Shetrone} M.,  {Venn} K.~A.,  {Tolstoy} E.,  {Primas} F.,  {Hill} V.,
  {Kaufer} A.,  2003, \mn@doi [\aj] {10.1086/345966}, \href
  {https://ui.adsabs.harvard.edu/abs/2003AJ....125..684S} {125, 684}

\bibitem[\protect\citeauthoryear{{Shipp} et~al.,}{{Shipp}
  et~al.}{2018}]{Shipp2018}
{Shipp} N.,  et~al., 2018, \mn@doi [\apj] {10.3847/1538-4357/aacdab}, \href
  {https://ui.adsabs.harvard.edu/abs/2018ApJ...862..114S} {862, 114}

\bibitem[\protect\citeauthoryear{{Simon}}{{Simon}}{2019}]{simon2019}
{Simon} J.~D.,  2019, \mn@doi [\araa] {10.1146/annurev-astro-091918-104453},
  \href {https://ui.adsabs.harvard.edu/abs/2019ARA&A..57..375S} {57, 375}

\bibitem[\protect\citeauthoryear{{Simon}, {Jacobson}, {Frebel}, {Thompson},
  {Adams}  \& {Shectman}}{{Simon} et~al.}{2015}]{Simon2015scl}
{Simon} J.~D.,  {Jacobson} H.~R.,  {Frebel} A.,  {Thompson} I.~B.,  {Adams}
  J.~J.,   {Shectman} S.~A.,  2015, \mn@doi [\apj]
  {10.1088/0004-637X/802/2/93}, \href
  {https://ui.adsabs.harvard.edu/abs/2015ApJ...802...93S} {802, 93}

\bibitem[\protect\citeauthoryear{{Sk{\'u}lad{\'o}ttir} \&
  {Salvadori}}{{Sk{\'u}lad{\'o}ttir} \&
  {Salvadori}}{2020}]{Skuladottir2020rprocess}
{Sk{\'u}lad{\'o}ttir} {\'A}.,  {Salvadori} S.,  2020, \mn@doi [\aap]
  {10.1051/0004-6361/201937293}, \href
  {https://ui.adsabs.harvard.edu/abs/2020A&A...634L...2S} {634, L2}

\bibitem[\protect\citeauthoryear{{Sk{\'u}lad{\'o}ttir}, {Tolstoy}, {Salvadori},
  {Hill}, {Pettini}, {Shetrone}  \& {Starkenburg}}{{Sk{\'u}lad{\'o}ttir}
  et~al.}{2015}]{Skuladottir2015cempScl}
{Sk{\'u}lad{\'o}ttir} {\'A}.,  {Tolstoy} E.,  {Salvadori} S.,  {Hill} V.,
  {Pettini} M.,  {Shetrone} M.~D.,   {Starkenburg} E.,  2015, \mn@doi [\aap]
  {10.1051/0004-6361/201424782}, \href
  {https://ui.adsabs.harvard.edu/abs/2015A&A...574A.129S} {574, A129}

\bibitem[\protect\citeauthoryear{{Sk{\'u}lad{\'o}ttir}, {Hansen}, {Salvadori}
  \& {Choplin}}{{Sk{\'u}lad{\'o}ttir} et~al.}{2019}]{Skuladottir2019}
{Sk{\'u}lad{\'o}ttir} {\'A}.,  {Hansen} C.~J.,  {Salvadori} S.,   {Choplin} A.,
   2019, \mn@doi [\aap] {10.1051/0004-6361/201936125}, \href
  {https://ui.adsabs.harvard.edu/abs/2019A&A...631A.171S} {631, A171}

\bibitem[\protect\citeauthoryear{{Sk{\'u}lad{\'o}ttir}
  et~al.,}{{Sk{\'u}lad{\'o}ttir} et~al.}{2021}]{Skuladottir2021UMPsculptor}
{Sk{\'u}lad{\'o}ttir} {\'A}.,  et~al., 2021, \mn@doi [\apjl]
  {10.3847/2041-8213/ac0dc2}, \href
  {https://ui.adsabs.harvard.edu/abs/2021ApJ...915L..30S} {915, L30}

\bibitem[\protect\citeauthoryear{{Sk{\'u}lad{\'o}ttir}, {Vanni}, {Salvadori}
  \& {Lucchesi}}{{Sk{\'u}lad{\'o}ttir} et~al.}{2023}]{Skuladottir2023Scl}
{Sk{\'u}lad{\'o}ttir} {\'A}.,  {Vanni} I.,  {Salvadori} S.,   {Lucchesi} R.,
  2023, \mn@doi [arXiv e-prints] {10.48550/arXiv.2305.02829}, \href
  {https://ui.adsabs.harvard.edu/abs/2023arXiv230502829S} {p. arXiv:2305.02829}

\bibitem[\protect\citeauthoryear{{Sneden}}{{Sneden}}{1973}]{Sneden1973moog}
{Sneden} C.~A.,  1973, PhD thesis, University of Texas, Austin

\bibitem[\protect\citeauthoryear{{Sneden}, {Cowan}  \& {Gallino}}{{Sneden}
  et~al.}{2008}]{Sneden2008}
{Sneden} C.,  {Cowan} J.~J.,   {Gallino} R.,  2008, \mn@doi [\araa]
  {10.1146/annurev.astro.46.060407.145207}, \href
  {https://ui.adsabs.harvard.edu/abs/2008ARA&A..46..241S} {46, 241}

\bibitem[\protect\citeauthoryear{{Sobeck} et~al.,}{{Sobeck}
  et~al.}{2011}]{Sobeck2011}
{Sobeck} J.~S.,  et~al., 2011, \mn@doi [\aj] {10.1088/0004-6256/141/6/175},
  \href {https://ui.adsabs.harvard.edu/abs/2011AJ....141..175S} {141, 175}

\bibitem[\protect\citeauthoryear{{Springel}, {Frenk}  \& {White}}{{Springel}
  et~al.}{2006}]{Springel2006}
{Springel} V.,  {Frenk} C.~S.,   {White} S. D.~M.,  2006, \mn@doi [\nat]
  {10.1038/nature04805}, \href
  {https://ui.adsabs.harvard.edu/abs/2006Natur.440.1137S} {440, 1137}

\bibitem[\protect\citeauthoryear{{Suda} et~al.,}{{Suda} et~al.}{2008}]{SAGA_1}
{Suda} T.,  et~al., 2008, \mn@doi [\pasj] {10.1093/pasj/60.5.1159}, \href
  {https://ui.adsabs.harvard.edu/abs/2008PASJ...60.1159S} {60, 1159}

\bibitem[\protect\citeauthoryear{{Suda} et~al.,}{{Suda} et~al.}{2017}]{SAGA_4}
{Suda} T.,  et~al., 2017, \mn@doi [\pasj] {10.1093/pasj/psx059}, \href
  {https://ui.adsabs.harvard.edu/abs/2017PASJ...69...76S} {69, 76}

\bibitem[\protect\citeauthoryear{{Tenachi}, {Oria}, {Ibata}, {Famaey}, {Yuan},
  {Arentsen}, {Martin}  \& {Viswanathan}}{{Tenachi}
  et~al.}{2022}]{Tenachi2022typhon}
{Tenachi} W.,  {Oria} P.-A.,  {Ibata} R.,  {Famaey} B.,  {Yuan} Z.,  {Arentsen}
  A.,  {Martin} N.,   {Viswanathan} A.,  2022, \mn@doi [\apjl]
  {10.3847/2041-8213/ac874f}, \href
  {https://ui.adsabs.harvard.edu/abs/2022ApJ...935L..22T} {935, L22}

\bibitem[\protect\citeauthoryear{{Ting}, {Conroy}, {Rix}  \& {Cargile}}{{Ting}
  et~al.}{2019}]{Ting2019ThePayne}
{Ting} Y.-S.,  {Conroy} C.,  {Rix} H.-W.,   {Cargile} P.,  2019, \mn@doi [\apj]
  {10.3847/1538-4357/ab2331}, \href
  {https://ui.adsabs.harvard.edu/abs/2019ApJ...879...69T} {879, 69}

\bibitem[\protect\citeauthoryear{{Tinsley}}{{Tinsley}}{1979}]{Tinsley1979}
{Tinsley} B.~M.,  1979, \mn@doi [\apj] {10.1086/157039}, \href
  {https://ui.adsabs.harvard.edu/abs/1979ApJ...229.1046T} {229, 1046}

\bibitem[\protect\citeauthoryear{{Tolstoy}, {Hill}  \& {Tosi}}{{Tolstoy}
  et~al.}{2009}]{Tolstoy2009}
{Tolstoy} E.,  {Hill} V.,   {Tosi} M.,  2009, \mn@doi [\araa]
  {10.1146/annurev-astro-082708-101650}, \href
  {https://ui.adsabs.harvard.edu/abs/2009ARA&A..47..371T} {47, 371}

\bibitem[\protect\citeauthoryear{{Travaglio}, {Gallino}, {Arnone}, {Cowan},
  {Jordan}  \& {Sneden}}{{Travaglio} et~al.}{2004}]{Travaglio2004sprocess}
{Travaglio} C.,  {Gallino} R.,  {Arnone} E.,  {Cowan} J.,  {Jordan} F.,
  {Sneden} C.,  2004, \mn@doi [\apj] {10.1086/380507}, \href
  {https://ui.adsabs.harvard.edu/abs/2004ApJ...601..864T} {601, 864}

\bibitem[\protect\citeauthoryear{{Tsujimoto}}{{Tsujimoto}}{2021}]{Tsujimoto2021rprocess}
{Tsujimoto} T.,  2021, \mn@doi [\apjl] {10.3847/2041-8213/ac2c75}, \href
  {https://ui.adsabs.harvard.edu/abs/2021ApJ...920L..32T} {920, L32}

\bibitem[\protect\citeauthoryear{{Tsujimoto}, {Ishigaki}, {Shigeyama}  \&
  {Aoki}}{{Tsujimoto} et~al.}{2015}]{Tsujimoto2015}
{Tsujimoto} T.,  {Ishigaki} M.~N.,  {Shigeyama} T.,   {Aoki} W.,  2015, \mn@doi
  [\pasj] {10.1093/pasj/psv035}, \href
  {https://ui.adsabs.harvard.edu/abs/2015PASJ...67L...3T} {67, L3}

\bibitem[\protect\citeauthoryear{{Usman} et~al.,}{{Usman}
  et~al.}{2024}]{Usman2024300s}
{Usman} S.~A.,  et~al., 2024, \mn@doi [\mnras] {10.1093/mnras/stae185}, \href
  {https://ui.adsabs.harvard.edu/abs/2024MNRAS.tmp..241U} {}

\bibitem[\protect\citeauthoryear{{Vasiliev}}{{Vasiliev}}{2019}]{agama}
{Vasiliev} E.,  2019, \mn@doi [\mnras] {10.1093/mnras/sty2672}, \href
  {https://ui.adsabs.harvard.edu/abs/2019MNRAS.482.1525V} {482, 1525}

\bibitem[\protect\citeauthoryear{{Vasiliev} \& {Baumgardt}}{{Vasiliev} \&
  {Baumgardt}}{2021}]{VasilievBaumgardt2021gcs}
{Vasiliev} E.,  {Baumgardt} H.,  2021, \mn@doi [\mnras]
  {10.1093/mnras/stab1475}, \href
  {https://ui.adsabs.harvard.edu/abs/2021MNRAS.505.5978V} {505, 5978}

\bibitem[\protect\citeauthoryear{{Wanajo}, {Hirai}  \& {Prantzos}}{{Wanajo}
  et~al.}{2021}]{Wanajo2021}
{Wanajo} S.,  {Hirai} Y.,   {Prantzos} N.,  2021, \mn@doi [\mnras]
  {10.1093/mnras/stab1655}, \href
  {https://ui.adsabs.harvard.edu/abs/2021MNRAS.505.5862W} {505, 5862}

\bibitem[\protect\citeauthoryear{{White} \& {Frenk}}{{White} \&
  {Frenk}}{1991}]{White1991cdm}
{White} S. D.~M.,  {Frenk} C.~S.,  1991, \mn@doi [\apj] {10.1086/170483}, \href
  {https://ui.adsabs.harvard.edu/abs/1991ApJ...379...52W} {379, 52}

\bibitem[\protect\citeauthoryear{{Yong} et~al.,}{{Yong}
  et~al.}{2013}]{yong2013full}
{Yong} D.,  et~al., 2013, \mn@doi [\apj] {10.1088/0004-637X/762/1/26}, \href
  {https://ui.adsabs.harvard.edu/abs/2013ApJ...762...26Y} {762, 26}

\bibitem[\protect\citeauthoryear{{Yoon} et~al.,}{{Yoon}
  et~al.}{2018}]{yoon2018}
{Yoon} J.,  et~al., 2018, \mn@doi [\apj] {10.3847/1538-4357/aaccea}, \href
  {https://ui.adsabs.harvard.edu/abs/2018ApJ...861..146Y} {861, 146}

\bibitem[\protect\citeauthoryear{{Yuan}, {Chang}, {Beers}  \& {Huang}}{{Yuan}
  et~al.}{2020}]{Yuan2020lms1}
{Yuan} Z.,  {Chang} J.,  {Beers} T.~C.,   {Huang} Y.,  2020, \mn@doi [\apjl]
  {10.3847/2041-8213/aba49f}, \href
  {https://ui.adsabs.harvard.edu/abs/2020ApJ...898L..37Y} {898, L37}

\bibitem[\protect\citeauthoryear{{de los Reyes}, {Kirby}, {Ji}  \&
  {Nu{\~n}ez}}{{de los Reyes} et~al.}{2022}]{delosReyes2022}
{de los Reyes} M. A.~C.,  {Kirby} E.~N.,  {Ji} A.~P.,   {Nu{\~n}ez} E.~H.,
  2022, \mn@doi [\apj] {10.3847/1538-4357/ac332b}, \href
  {https://ui.adsabs.harvard.edu/abs/2022ApJ...925...66D} {925, 66}

\makeatother
\end{thebibliography}

\bsp	
\label{lastpage}
\end{document}